\documentclass[aps, prx, twocolumn, superscriptaddress]{revtex4-2}

\usepackage{hyperref, graphicx, xcolor, amsmath, amssymb, siunitx, ulem}
\usepackage[caption=false]{subfig}
\usepackage{verbatim}
\usepackage{comment}
\usepackage{braket}
\usepackage{etoolbox}
\usepackage{physics}

\usepackage{natbib}
\hypersetup{colorlinks, linkcolor=teal, citecolor=blue, urlcolor=purple}
\newrobustcmd*{\citefirstlastauthor}{\AtNextCite{\DeclareNameAlias{labelname}{given-family}}\citeauthor}

\newcommand{\kket}[1]{|{#1}\rangle\rangle}

\newcommand{\ba}{\mathbf{a}}
\newcommand{\bb}{\mathbf{b}}
\newcommand{\bp}{\boldsymbol{\varphi}}
\newcommand{\bD}{\mathbf{D}}
\newcommand{\bP}{\mathbf{P}}
\newcommand{\bH}{\mathbf{H}}
\newcommand{\bL}{\mathbf{L}}
\newcommand{\bZ}{\mathbf{Z}}
\newcommand{\bX}{\mathbf{X}}
\newcommand{\bY}{\mathbf{Y}}
\newcommand{\bG}{\mathbf{G}}
\newcommand{\brho}{\boldsymbol{\rho}}

\makeatletter
\def\@fnsymbol#1{\ensuremath{\ifcase#1\or \dagger\or \ddagger\or
   \mathsection\or \mathparagraph\or \|\or **\or \dagger\dagger
   \or \ddagger\ddagger \else\@ctrerr\fi}}
\makeatother

\newcommand{\AB}{Alice \& Bob, 53 Bd du Général Martial Valin, 75015 Paris, France}

\newcommand{\ENSL}{Ecole Normale Supérieure de Lyon, CNRS, Laboratoire de Physique, F-69342 Lyon, France}

\newcommand{\LPENS}{Laboratoire de Physique de l'Ecole normale supérieure, ENS-PSL, CNRS, Sorbonne Université, Université Paris Cité, Centre Automatique et Systèmes, Mines Paris, Université PSL, Inria, Paris, France}

\begin{document}

\title{Quantum control of a cat-qubit with bit-flip times exceeding ten seconds}

\author{U.~Réglade}
\thanks{These authors contributed equally to the project}
\affiliation{\AB}
\affiliation{\LPENS}

\author{A.~Bocquet}
\thanks{These authors contributed equally to the project}
\affiliation{\AB}
\affiliation{\LPENS}
\author{R.~Gautier}
\affiliation{\LPENS}
\author{J.~Cohen}
\affiliation{\AB}
\author{A.~Marquet}
\affiliation{\AB}
\affiliation{\ENSL}
\author{E.\ Albertinale}
\affiliation{\AB}
\author{N.\ Pankratova}
\affiliation{\AB}
\author{M.\ Hallén}
\affiliation{\AB}
\author{F.\ Rautschke}
\affiliation{\AB}
\author{L.-A.~Sellem}
\affiliation{\LPENS}
\author{P.~Rouchon}
\affiliation{\LPENS}
\author{A.~Sarlette}
\affiliation{\LPENS}
\author{M.~Mirrahimi}
\affiliation{\LPENS}
\author{P.~Campagne-Ibarcq}
\affiliation{\LPENS}
\author{R.~Lescanne}
\affiliation{\AB}
\author{S.~Jezouin}
\thanks{These authors co-supervised the project}
\affiliation{\AB}
\author{Z.~Leghtas}
\thanks{These authors co-supervised the project}
\affiliation{\LPENS}


\begin{abstract}
Quantum bits (qubits) are prone to several types of errors due to uncontrolled interactions with their environment. Common strategies to correct these errors are based on architectures of qubits involving daunting hardware overheads \cite{Google2023}. A hopeful path forward is to build qubits that are inherently protected against certain types of errors, so that the overhead required to correct remaining ones is significantly reduced  \cite{Aliferis2008, Webster2015, Tuckett2018, Guillaud2019, Darmawan2021, ruiz2024}. However, the foreseen benefit rests on a severe condition: quantum manipulations of the qubit must not break the protection that has been so carefully engineered \cite{Guillaud2019, Puri2020}. A recent qubit - the cat-qubit - is encoded in the manifold of metastable states of a quantum dynamical system, thereby acquiring continuous and autonomous protection against bit-flips. Here, in a superconducting circuit experiment, we implement a cat-qubit with bit-flip times exceeding 10 seconds. This is a four order of magnitude improvement over previous cat-qubit implementations. We prepare and image quantum superposition states, and measure phase-flip times above 490 nanoseconds. Most importantly, we control the phase of these quantum superpositions without breaking bit-flip protection. This experiment demonstrates the compatibility of quantum control and inherent bit-flip protection at an unprecedented level, showing the viability of these dynamical qubits for future quantum technologies.

\end{abstract}

\maketitle

A dynamical system stems from the interplay of external forces, nonlinearities and dissipation \cite{Guckenheimer1983}. Of particular interest are bistable dynamical systems that switch between two attractors, such as the reversal of earth's magnetic field. At a vastly reduced scale, driven nonlinear oscillators containing only a handful of photons have displayed switching times of several seconds \cite{Muppalla2018}, making them ideal candidates for ultra-low power classical logic processing \cite{Mabuchi2011}.

It is then tempting to benefit from this stability to robustly encode quantum information, where susceptibility to noise continues to be the limiting factor for the emergence of quantum machines \cite{Google2023}. Qubits fail in two ways. First, the random switching between computational states - bit-flips - and second, the scrambling of the phase of quantum superpositions - phase-flips \cite{Nielsen2010}. A qubit encoded in the manifold of metastable states of a dynamical system, coined the cat-qubit, would be protected against bit-flips at the hardware level. The challenge is then to measure and control this qubit without breaking its protection. If this challenge is met, the only remaining error - phase-flips - can then be corrected by embedding these qubits in error-correcting architectures with substantially reduced hardware overhead \cite{Aliferis2008, Tuckett2018, Puri2020, Darmawan2021, Chamberland2022} in comparison with those required to correct both bit-flips and phase-flips \cite{Fowler2012, Google2023}.

{Making the leap from classical to quantum information processing with dynamical bistable systems is difficult}. Indeed, they owe their stability to friction - or dissipation - that dampens erroneous diffusion between states. However, friction commonly originates from interactions with an ensemble of degrees of freedom. This leaks information about the system, and quantum superpositions decohere into classical mixtures \cite{Zurek2003}. Surprisingly, there exists a type of dissipation, known as two-photon dissipation \cite{Wolinsky1988, Leghtas2015, Mirrahimi2014}, that provides stability without inducing decoherence. Indeed, two-photon exchanges between an oscillator and a cold environment is expected to stabilize two coherent states with macroscopic bit-flip times, while permitting the preparation and manipulation of their quantum superpositions \cite{Mirrahimi2014}. 

In practice, two-photon dissipation is implemented in a superconducting oscillator mode - the memory - that is coupled to a lossy buffer mode through a non linear Josephson element. In previous experiments, quantum tomography of the memory was performed via an ancillary system composed of a transmon and its readout resonator. While quantum superpositions of two metastable states were observed, the bit-flip time saturated in the millisecond range \cite{Lescanne2020}. Cat-qubit implementations based on the Kerr effect reached similar timescales \cite{Grimm2020, Frattini2022}. In a recent experiment \cite{Berdou2023}, this tomography apparatus was entirely removed and bit-flip times exceeding one hundred seconds were observed. However, since the two-photon exchange rate was dominated by single-photon loss, quantum superposition states could not be prepared nor measured, thereby falling short from implementing a qubit. This incriminating evidence motivated the removal of the ancillary transmon and the development of an alternative tomography procedure that does not break bit-flip protection.

\begin{figure}
\centering
\includegraphics[width=0.85\columnwidth]{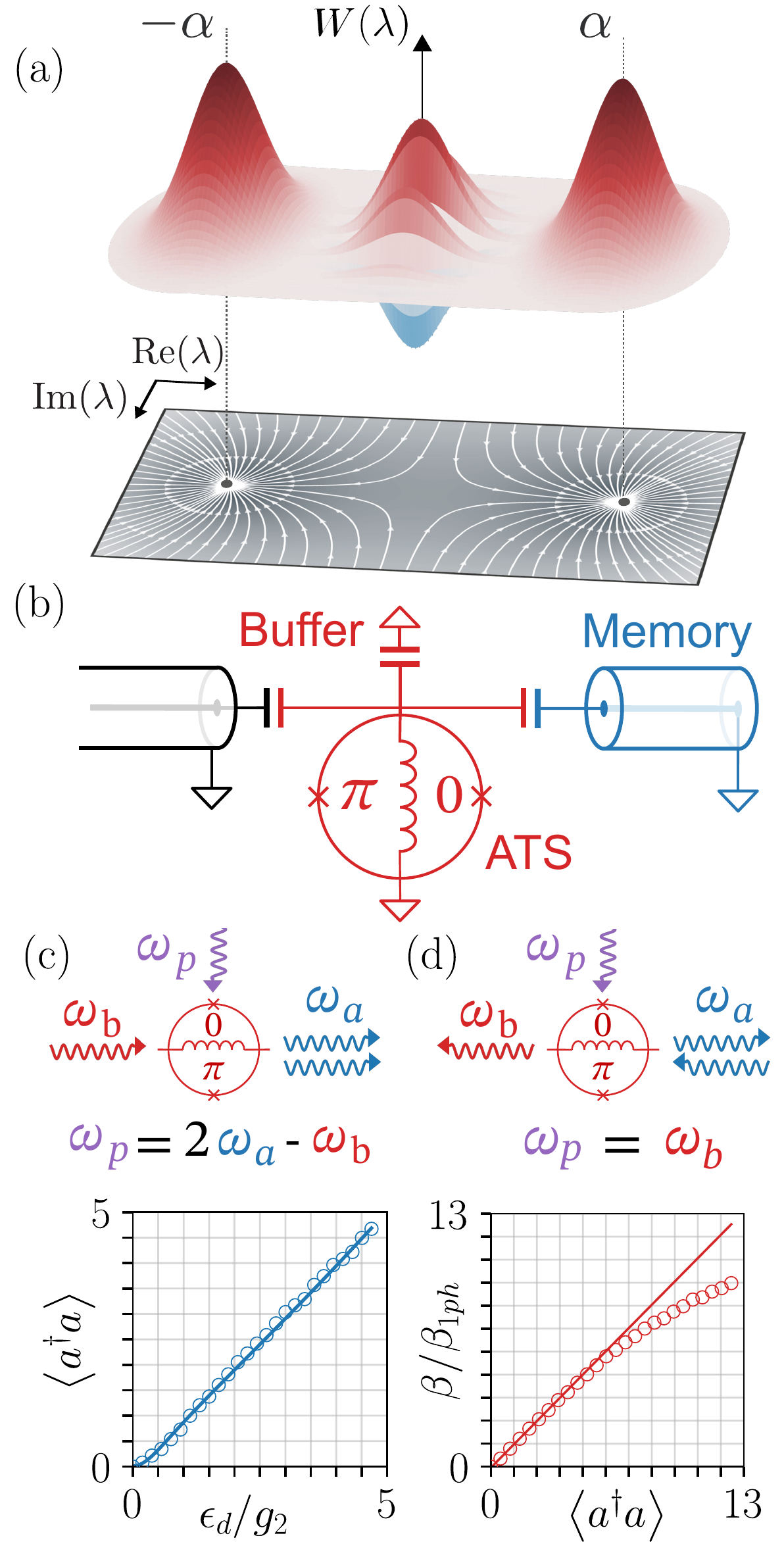}
\caption{{\bf Encoding quantum information in a bistable dynamical system.} (a) Semi-classical trajectories (solid lines, derived from Eq.~\eqref{eq:H}) converge towards two metastable states. Quantum information encoded in the manifold spanned by these states (the Wigner function corresponds to a coherent superposition state) inherit a protection against bit-flips. (b) Circuit implementation of our dynamical system. A quarter wavelength transmission line mode (memory in blue) is coupled to its environment (black) through a buffer mode (red) composed of an asymmetrically threaded SQUID (ATS) (see text). (c) A pump (purple) and a buffer drive (red) combine through the ATS to inject photon pairs in the memory (blue). The reverse process that removes photon pairs is not represented. We measure (open circles) a linear increase (simulation in solid line) of the steady-state memory photon number (y-axis) versus drive amplitude (x-axis) (d) A pump at frequency $\omega_b$ (purple) displaces the buffer (red) conditionally on the number of photons in the memory (blue), a key ingredient for our quantum tomography protocol. We measure (open circles) the buffer amplitude (y-axis) versus the memory photon number (x-axis) reached after a displacement pulse. The deviation from the linear trend (solid line) is a signature of compression due to higher order processes.}
\label{fig:fig1}
\end{figure}

In this experiment, we implement a cat-qubit with bit-flip times exceeding ten seconds, an improvement of four orders of magnitude over previous cat-qubit implementations, and six orders of magnitude over the lifetime of the photons composing the qubit. We observe phase-flip times greater than 490~ns, mainly limited by single-photon loss. We control the phase of coherent superpositions by rotating in a Zeno-blocked manifold \cite{Touzard2018}, performing a $\pi$ rotation around $Z$ in 235~ns. We verify that this manipulation only marginally reduces the bit-flip time, maintaining it above ten seconds. This was made possible by implementing a quantum tomography protocol that requires no additional ancillary elements \cite{Albert2016}. Indeed, the Josephson dipole that mediates two-photon dissipation is operated to map quantum observables of the memory onto the buffer. This experiment demonstrates the tomography and control of a cat-qubit without breaking bit-flip protection at the level of bit-flip times of ten seconds. However, further improvements in state preparation, measurement fidelities and single photon loss will be necessary before scaling to a fully-protected hardware-efficient logical qubit \cite{Guillaud2019, Puri2020, Darmawan2021, Chamberland2022}.

Our dynamical system is well described by the following Hamiltonian and loss operator:
\begin{eqnarray}
    \bH_{2ph}&=&{g_2^*}\ba^2\bb^\dag+g_2{\ba^\dag}^2\bb - \varepsilon_d^*\bb-\varepsilon_d\bb^\dag\;,\notag\\
    \bL_b&=&\sqrt{\kappa_b}\bb\;,\label{eq:H}
\end{eqnarray}
where $\ba, \bb$ are respectively the memory and buffer annihilation operators, $\varepsilon_d$ the amplitude of a resonant drive applied to the buffer, and $\kappa_{b}$ the buffer energy damping rate. Photon pairs are dissipated from the memory by converting them at rate $g_2$ to single photons in the buffer, which are then dissipated into the environment. In the absence of energy damping in the memory, the steady states of this system lie in a two-dimensional manifold \cite{Leghtas2015} spanned by{ 
\begin{equation}
\ket{\pm}_\alpha = (\ket{\alpha}\pm\ket{-\alpha})/\sqrt{\mathcal{N}_\pm}\;,
\end{equation}
where $ \mathcal{N}_\pm={2 \pm 2\exp(-2\left|\alpha\right|^2)}$ and $\ket{\alpha}$ is a coherent state of amplitude $\alpha$ that is controlled by the drive amplitude: $\alpha^2=\varepsilon_d/g_2^*$}. The local convergence rate towards this manifold is denoted $\kappa_\text{conf}$ and in our parameter regime it saturates at $\kappa_\text{conf}\approx\kappa_b/2$ (see Supplementary Sec.~\ref{sec:confinement_rate}). The qubit encoded in this manifold owes its name - the cat-qubit \cite{Mirrahimi2014} - to the fact that $\ket{\pm}_{\alpha}$ resemble Schr\"odinger cat states for $|\alpha|\gtrsim 1$ \cite{Haroche2006}. Its computational states are defined as $\ket{0/1}_\alpha=(\ket{+}_\alpha\pm\ket{-}_\alpha)/\sqrt{2}\approx\ket{\pm\alpha}$, and its $X$ and $Z$ Pauli operator as $\bZ_\alpha\approx\ketbra{\alpha}{\alpha}-\ketbra{-\alpha}{-\alpha}$ and $\bX_\alpha\approx\ketbra{\alpha}{-\alpha}+\ketbra{-\alpha}{\alpha}$ up to errors that are exponentially small in $|\alpha|^2$.

\begin{figure*}
\centering
\includegraphics[width=2.15\columnwidth]{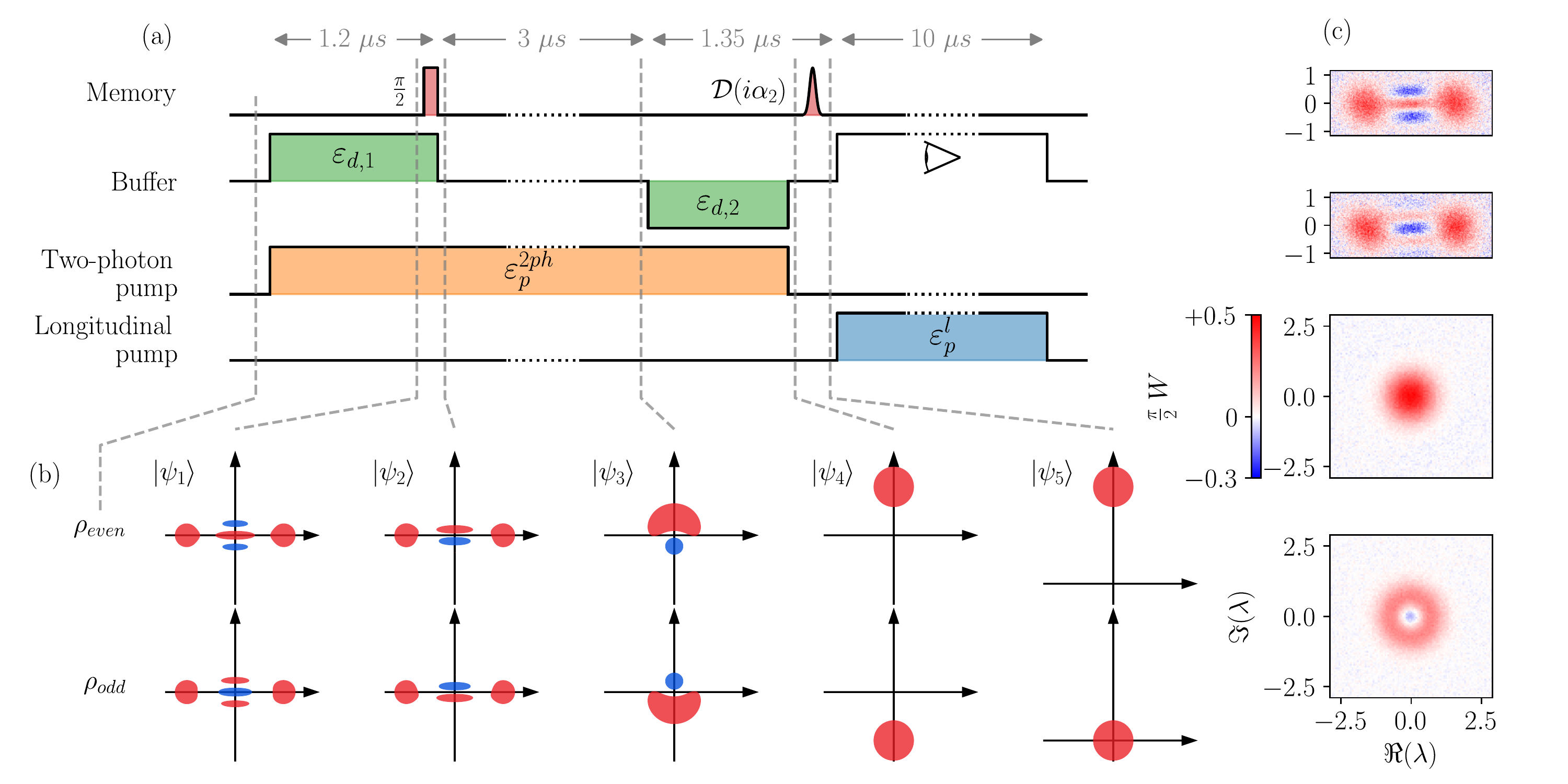}
\caption{{\bf Quantum tomography protocol based on the holonomic gate} \cite{Albert2016}. (a) Pulse sequence for each control channel as a function of time. The parity of the memory is mapped onto photon number, which is measured through longitudinal coupling to the buffer (see text). (b) Cartoon of the memory Wigner function at each step (dashed lines) of the protocol for an even (top) and odd (bottom) initial state. (c) Measured Wigner functions of the memory initialized in : $\ket{+}_\alpha, \ket{-}_\alpha, \ket{0}$ and Fock state $\ket{1}$ (top to bottom), obtained by combining the photon parity measurement with simple memory displacements and an active memory reset (see Supplementary Sec.~\ref{sec:reset}). The first two images contain $250\times 100$ pixels averaged 5000 times, and the last two contain $100\times 100$ pixels averaged 70000 times. Their acquisition time is 2 and 12 hours respectively.}
\label{fig:fig2}
\end{figure*}

\begin{figure*}
\centering
\includegraphics[width=2.0\columnwidth]{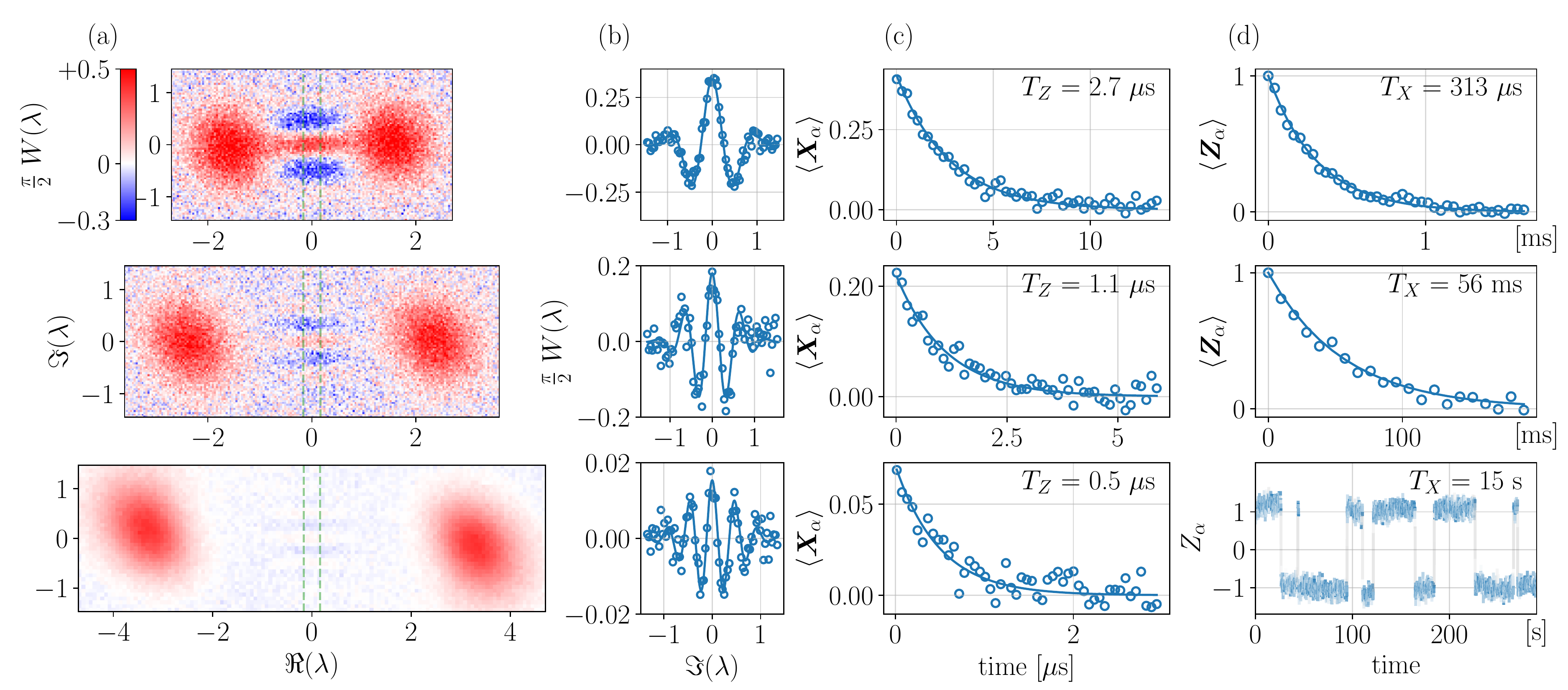}
\caption{{\bf Cat-qubit phase-flip and bit-flip time measurements.} Each row of figures represents a photon number $|\alpha|^2=2.5, 5.6, 11.3$ (top, middle, bottom). (a) Measured Wigner functions of the memory state $\ket{+}_\alpha$ prepared in $1~\mu$s. Constraining these maps to sum to one sets an absolute scale for our parity measurement. (b) Integration of the measured Wigner functions (y-axis) over the window delimited by the green dashed lines in (a) versus the imaginary axis (x-axis). We fit the analytical formula of these oscillations \cite{Haroche2006} (solid lines) to the data (open circles). (c) Evolution of the expectation value of $\bX_\alpha$ (y-axis) versus time (x-axis) for a memory state $\ket{+}_\alpha$ prepared in $400$~ns. The data (open circles) follow an exponential decay (blue solid line) from which we extract the phase-flip time $T_Z$. (d-top,middle) Expectation value of $\bZ_\alpha$ (y-axis) versus time (x-axis). The data (open circles) follow an exponential decay (solid line) from which we extract the bit-flip time $T_X$. (d-bottom) Real-time trajectory cropped from the full data-set of the memory switching between $Z_\alpha=\pm 1$ (y-axis) versus time (x-axis). It is acquired by applying a drive on the memory and collecting the buffer fluorescence (solid line) (see Supplementary Sec.~\ref{sec:memory_trajectories}).}
\label{fig:fig3}
\end{figure*}

{States $\ket{\pm\alpha}$ are localized in opposite sides of phase-space (Fig.~1a), with exponentially small support overlap in $|\alpha|^2$}. Therefore, even in the inevitable presence of losses, provided they are diffusive-like \cite{Gottesman2001} and weak compared to $\kappa_\text{conf}$, the bit-flip time $T_X$ between $\ket{\pm\alpha}$ is expected to increase exponentially with $|\alpha|^2$ \cite{Lescanne2020}. From $|\alpha|^2\approx 10$ onwards, timescales $T_X$ exceeding seconds are predicted in our parameter regime. Quantum superpositions of $\ket{\pm\alpha}$ are prepared by initializing the memory in the vacuum, and activating the two-photon exchange mechanism \cite{Leghtas2015}. Since the dynamics of Eq.~\eqref{eq:H} conserve memory photon number parity, the state spontaneously dissipates towards $\ket{+}_\alpha$ on a timescale set by $\kappa_\text{conf}^{-1}$. The state then evolves into a classical statistical mixture of $\ket{\pm\alpha}$ at rate $\Gamma^{\kappa_a}_Z=2\kappa_a|\alpha|^2$ \cite{Haroche2006}, where $\kappa_a$ is the memory energy damping rate.  Therefore, the observation of quantum superpositions of metastable states with macroscopic bit-flip times requires that the decoherence rate  verifies $\Gamma^{\kappa_a}_Z<\kappa_\text{conf}$ for $|\alpha^2|\approx 10$.

We implement the dynamics of Eq.~\eqref{eq:H} in a two-dimensional circuit quantum electrodynamics (cQED) architecture (Fig.~1b) \cite{Girvin2014} operated in a dilution refrigerator at 10~mK. The chip consists of a sapphire substrate on which we sputter a tantalum film \cite{Place2021}, which is then patterned. The memory is a quarter wavelength coplanar waveguide resonator of frequency $\omega_a/2\pi=5.26$~GHz and decay rate $\kappa_a/2\pi=9.3$~kHz, corresponding to a lifetime of 17~$\mu$s. It is capacitively coupled to the buffer: an island shunted to ground through a nonlinear element called the ATS (Asymmetrically Threaded SQUID) \cite{Lescanne2020}, resonating at $\omega_b/2\pi=7.70$~GHz with decay rate $\kappa_b/2\pi=2.6$~MHz. The ATS is composed of a SQUID (Superconducting Quantum Interference Device) shunted in the middle by a kinetic inductance thus forming two loops. By setting the flux to 0 and $\pi$ in the right and left loops respectively (Fig.~1b), this element induces the following nonlinear potential $U(\bp)\approx-2E_J\epsilon_p(t)\sin(\bp)$ \cite{Lescanne2020}, where $E_J$ is the Josephson energy of the SQUID junctions. Here, $\epsilon_p(t)=\varepsilon_p\cos(\omega_p t)$ is a flux-pump of amplitude $\varepsilon_p$ and angular frequency $\omega_p$, and $\bp$ is the phase drop across the ATS which is a linear combination of $\ba, \ba^\dag, \bb, \bb^\dag$. Setting the pump frequency to $\omega_p=2\omega_a-\omega_b$ activates the desired third order process ${\ba}^2\bb^\dag+{\ba^\dag}^2\bb$ (Fig.~1c) at a rate $g_2$ that grows linearly with the pump amplitude. The latter is increased until $g_2$ is on par with $\kappa_b$, thereby maximizing $\kappa_\text{conf}$. We reach $g_2/2\pi=0.76 $~MHz and $\kappa_\text{conf}/2\pi \approx 1.3$~MHz. For $|\alpha|^2 \in [1.4, 11.3]$, this places us in the favorable regime where $\Gamma^{\kappa_a}_Z/\kappa_\text{conf} \in [0.02, 0.16]$.

Josephson circuits have been referred to as the Swiss army knife of microwave quantum optics \cite{Flurin2014}. Simply by switching the pump frequency, the behavior of a dipole can be dramatically changed. By setting the pump frequency to $\omega_p = \omega_b$ (Fig.~1d), the following low order processes are resonantly selected: $(\bb+\bb^\dag)$, $\ba^\dag \ba (\bb+\bb^\dag)$ and ${\bb^\dag}^2 \bb+\bb^\dag\bb^2$. The first term $(\bb+\bb^\dag)$ is canceled by adding an additional drive of equal amplitude and opposite phase on the buffer. The second term $\ba^\dag \ba (\bb+\bb^\dag)$ is analogous to the radiation pressure coupling in optomechanics \cite{aspelmeyer2014cavity}, and has been referred to as a ``longitudinal" coupling in the context of Josephson circuits \cite{Touzard2019}. Conditioned on the number of photons $n_a$ in the memory, the buffer converges towards a coherent state of amplitude denoted $\beta_{1ph}\times n_a$. When cascaded with a heterodyne detection of the buffer, it constitutes a quantum non demolition (QND) measurement of the memory photon number. The third term ${\bb^\dag}^2 \bb+\bb^\dag\bb^2$ on the other hand is a parasitic interaction that is responsible for the compression visible in Fig.~1d thereby limiting the dynamical range of our detector. The longitudinal pump amplitude is chosen to maximize the detection efficiency over a dynamical range of 0 to about 10 photons in the memory. We reach a single shot fidelity of $89\%$ to distinguish between the vacuum and a coherent state containing 10 photons with an integration time of $10~\mu$s constrained by the memory lifetime (see Supplementary Sec.~\ref{sec:longitudinal_coupling}).

\begin{figure}
\centering
\includegraphics[width=1\columnwidth]{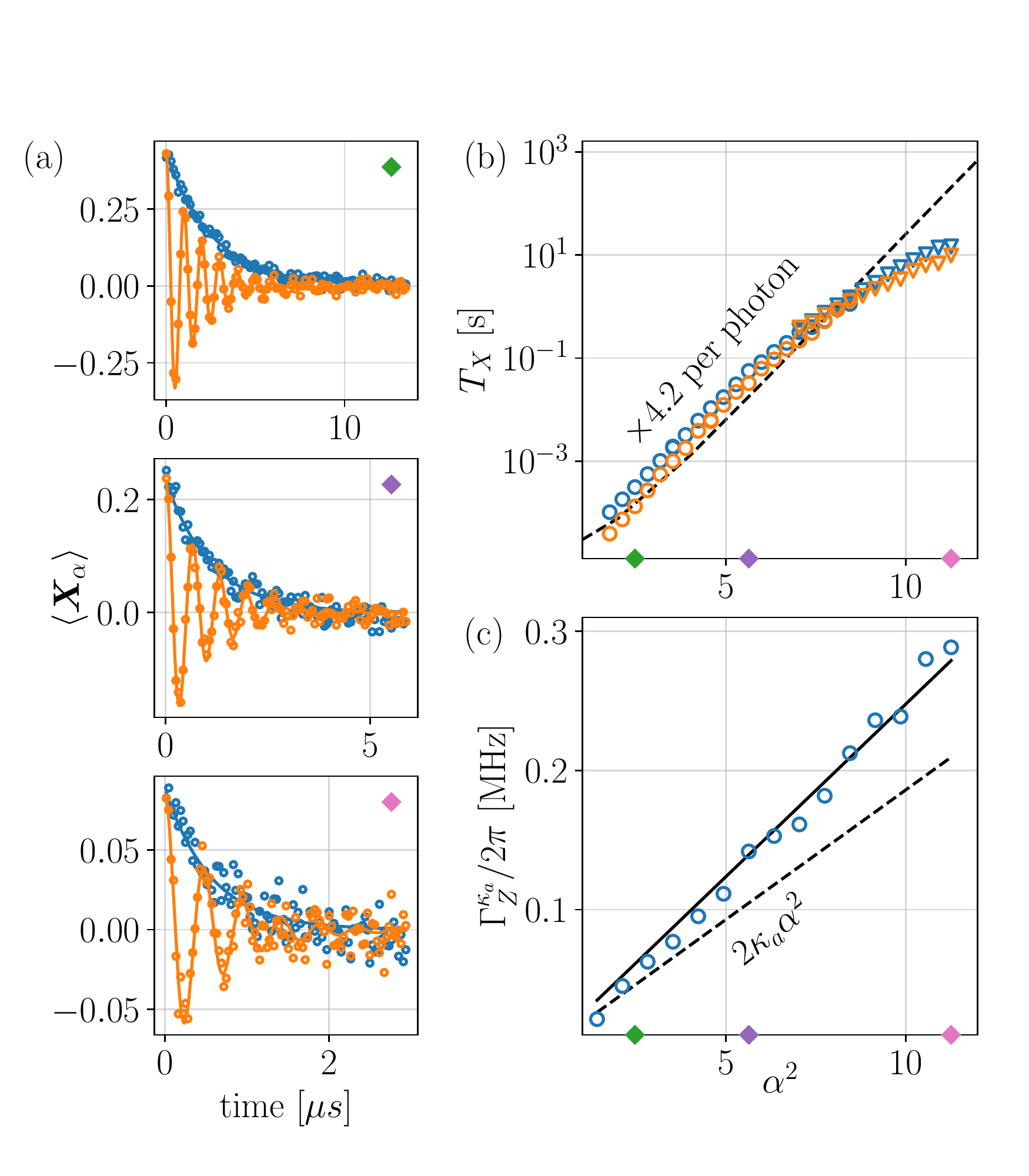} 
\caption{{\bf Quantum control that preserves bit-flip protection.} (a) The cat-qubit is initialized in $\ket{+}_\alpha$ for $|\alpha^2|=2.5, 5.6, 11.3$ (top, middle, bottom) with a preparation duration of 400~ns, and the expectation value of $\bX_\alpha$ (y-axis) is measured (open circles) versus time (x-axis). In the presence of a memory drive (orange), the cat-qubit undergoes coherent Zeno-blocked oscillations around its $Z$ axis. These oscillations decay exponentially with time. The decay in absence of oscillations (blue) are superimposed for reference. (b) The bit-flip time (y-axis, log scale) increases exponentially with photon number (x-axis), multiplying by 4.2 for each added photon, up to about 7 photons. It is extracted from the measured (open circles) or simulated (dashed line) average population transfer over time between $\ket{\pm\alpha}$, or from measured single real-time trajectories (triangles). (c) Dephasing rate (y-axis) as a function of photon number (x-axis). It follows a linear trend (solid line) with a slope that is $25\%$ larger than the one given by memory dissipation rate $\kappa_a$ (dashed line). The data points (circles) corresponding to panels (a) are marked in full dots. The data in (b,c) is taken in the absence (blue) or presence (orange) of a memory drive of same amplitude as in (a).}
\label{fig:fig4}
\end{figure}

We witness the quantum nature of the memory field through Wigner tomography \cite{Haroche2006}. The Wigner function is a quasi-probability distribution defined over the complex plane as $W(\lambda)=\frac{2}{\pi}\langle{\bD_\lambda \bP \bD_{-\lambda}}\rangle$, which represents the normalized expectation value of the parity operator $\bP=\exp(i\pi \ba^\dag \ba)$ for the state displaced by $\bD_\lambda=\exp(\lambda \ba^\dag-\lambda^*\ba)$. This graphical representation can display negativities which unambiguously testify of the non-classical nature of the field state. 

Our Wigner tomography protocol (Fig.~2) is based on the so-called holonomic gate proposed by Albert {\it et. al} \cite{Albert2016}. All odd parity states are mapped onto the vacuum $\ket{0}$, and all even parity states onto a coherent state $\ket{\psi_5}=\ket{2i\alpha_2}$ where $2\alpha_2\approx 4.8$. The QND photon number measurement through longitudinal coupling then distinguishes between $\ket{0}$ and $\ket{2i\alpha_2}$, providing a photon number parity measurement. Our pulse sequence (Fig.~2a) alternates between memory drives, two-photon dissipation, buffer drives of various amplitudes and longitudinal coupling. We now describe it step by step (Fig.~2b). Let us assume that the memory is initially in an even parity state $\rho_\text{even}$. First, we activate a two-photon pump $\varepsilon_p^{2ph}$ and buffer drive $\varepsilon_{d,1}=g_2^*\alpha_1^2$ where $\alpha_1$ is real. By parity conservation, $\rho_\text{even}$ is mapped to the even state $\ket{\psi_1}=\ket{+}_{\alpha_1}$. Next, we add a memory drive along the imaginary axis. Two-photon dissipation confines the dynamics to the quantum manifold spanned by $\ket{\pm}_{\alpha_1}$. The added memory drive induces coherent Zeno-blocked oscillations around the cat-qubit $Z$ axis \cite{Touzard2018}. We tune the drive length to perform a $\pi/2$ rotation reaching the parity-less state  $\ket{\psi_2}=(\ket{+}_{\alpha_1}+i\ket{-}_{\alpha_1})/\sqrt{2}$ \cite{Yurke1998, Kirchmair2013, Grimm2020}. Next, we turn off the memory and buffer drives while the two-photon pump remains active, thereby removing pairs of photons from $\ket{\psi_2}$. By parity conservation, $\ket{+}_{\alpha_1}$ is mapped to $\ket{0}$ and $\ket{-}_{\alpha_1}$ to $\ket{1}$ \cite{Grimm2020}. When this mapping is adiabatic with respect to $\kappa_\text{conf}^{-1}$ and fast compared to $\kappa_a^{-1}$, quantum superpositions are preserved, yielding $\ket{\psi_3}=(\ket{0}+i\ket{1})/\sqrt{2}$ \cite{Albert2016}. In practice we use a square buffer drive of amplitude $\alpha_1\approx1.6$ in an attempt to maximize the fidelity of the $\pi/2$ gate around $Z$, while minimizing the loss of coherence during these mappings. Next, maintaining the two-photon pump, we activate a buffer drive $\varepsilon_{d,2}=-g_2^*\alpha_2^2$ where  $\alpha_2$ is real. The minus sign on the buffer drive translates into a pure imaginary amplitude for the stabilized coherent states. This maps $\ket{0}\rightarrow\ket{+}_{i\alpha_2}$, $\ket{1}\rightarrow -i\ket{-}_{i\alpha_2}$ \cite{Grimm2020} and following the same reasoning: $\ket{\psi_3}\rightarrow \ket{\psi_4}=\ket{i\alpha_2}$ 
 \cite{Albert2016}. Conversely, an odd parity $\rho_\text{odd}$ would be mapped to $\ket{-i\alpha_2}$. Information on the parity of $\rho_\text{even/odd}$ is now encoded on the amplitude of coherent states $\ket{\pm i\alpha_2}$. Finally, the two-photon pump is turned off, and the memory is displaced by $i\alpha_2$. The longitudinal pump $\varepsilon_p^l$ is activated to distinguish between 0 and $4|\alpha_2|^2$ photons in the memory by heterodyne detection of the buffer. Note that the value of $\alpha_2$ can be tuned to optimize the fidelity of the longitudinal readout. Preceding this entire sequence by a memory displacement of amplitude $\lambda$ therefore measures $W(\lambda)$. We demonstrate this tomography protocol by measuring the Wigner functions of the vacuum $\ket{0}$, Fock state $\ket{1}$, and $\ket{\pm}_\alpha$ (Fig.~2c). The vacuum is prepared simply by waiting for several $\kappa_a^{-1}$ for the memory to settle in its thermodynamic equilibrium. Preparing $\ket{+}_\alpha$ requires the activation of a two-photon pump and buffer drive for several $\kappa_\text{conf}^{-1}$. Preparing $\ket{-}_\alpha$ requires an additional memory drive to perform a full Zeno-blocked $\pi$ rotation. Finally, from this state, switching
off the buffer drive while the two-photon pump remains
active prepares Fock state $\ket{1}$ by parity conservation.

The measurements of phase-flip and bit-flip times of our cat-qubit are displayed in Fig.~3. We prepare $\ket{+}_\alpha$ for various average photon numbers $|\alpha|^2$ by starting from a memory mode in the vacuum, and activating the corresponding buffer drive amplitude and two-photon pump. The preparation duration (400~ns or 1~$\mu$s, see Fig.~3), is chosen to be longer than $1/\kappa_\text{conf}\approx 120$~ns, ensuring sufficient time to reach the steady state manifold, and on par with $T_Z\approx 490$~ns for the largest states at 11.3 photons, ensuring the preservation of measurable quantum coherence. Using our novel tomography tool, we image the Wigner functions of these states and observe interference fringes that take negative values. While the contrast of these fringes reduces with $|\alpha|^2$, they remain visible up to $|\alpha|^2\approx 11.3$ photons (Fig.~3a,b). Note that in the cat-qubit code-space $\langle{\bX_\alpha}\rangle=\langle{\bP}\rangle$, and hence we extract the phase-flip time by monitoring the photon number parity decay over time. We measure phase-flip times ranging from $T_Z = 2.7~\mu$s for $|\alpha|^2=2.5$ to $T_Z = 490~$ns for $|\alpha|^2=11.3$ (Fig.~3c). Finally, we monitor the switching between $\ket{\pm\alpha}$ over time (Fig.~3d). To this end we prepare $\ket{+\alpha}$ by displacing the memory from the vacuum, before applying the two-photon pump and a buffer drive whose amplitude is adjusted to stabilize $\ket{\pm\alpha}$ for a variable time $t$. During this time, the state may switch to $\ket{-\alpha}$, causing a bit-flip. We detect the population of $\ket{\alpha}$ at time $t$ by setting the buffer drive to map $\ket{\pm\alpha}\rightarrow \ket{\pm\alpha'}$ where $\alpha'\approx2.1$, then interrupting the pump and buffer drive, and finally displacing the memory by $\alpha'$. This maps $\ket{-\alpha'}\rightarrow \ket{0}$ and $\ket{\alpha'}\rightarrow \ket{2\alpha'}$. Next, we activate the longitudinal pump to distinguish between these two states. For bit-flip times exceeding $\approx 100$~ms, this method leads to impractically long acquisition times. Instead, for these long bit-flip times that occur at $|\alpha|^2\gtrsim 7$, we sample the real-time trajectory of the memory field. After initializing the memory in $\ket{\alpha}$ and activating the two-photon exchange, we apply a weak drive of amplitude $\varepsilon_Z$ on the memory for 250~$\mu$s  every millisecond. This slightly displaces the state out of the steady state manifold. In response to this perturbation, the buffer develops an average field amplitude $\langle \bb \rangle=\mp \frac{\varepsilon_Z}{2\alpha^*g_2}$ depending on the state $\ket{\pm\alpha}$ in the memory \cite{Gautier2023} (see Supplementary  Sec.~\ref{sec:memory_trajectories
}). This field is then integrated by heterodyne detection (Fig.~3d, bottom panel) for the pulse duration of $T_\text{int}=250~\mu$s. For $\alpha^2\gtrsim7$, we have $T_Z\ll T_\text{int}\ll T_X$ and hence we observe bit-flip events in real time, so $T_X$ is well estimated from a single trace lasting $\approx 100~T_X$. Using these methods, we measure $T_X$ for $|\alpha|^2=2.5,5.6,11.3$, and observe a spectacular increase from $313~\mu$s, to $56$~ms, to 15 seconds.

We demonstrate quantum control of our cat-qubit and its impact on bit-flip protection in Fig.~4. After preparing the cat-qubit in $\ket{+}_\alpha$, we add a drive of amplitude $\varepsilon_Z$ on the memory mode. The interplay of this coherent drive and two-photon dissipation induces Zeno-blocked oscillations \cite{Touzard2018} at angular-frequency $\Omega_Z=4|\alpha|\varepsilon_Z$ \cite{Mirrahimi2014}, together with a drive-induced dephasing that increases with $\varepsilon_Z$ (see Supplementary  Sec.~\ref{sec:gammaZ_epsilonZ}). For $|\alpha^2|=11.3$, we observe a $\pi$ rotation in 235~ns. Since the benefit of the cat-qubit is to lighten the hardware overhead for error correction by sparing the need for active bit-flip correction, it is crucial to verify that our quantum manipulations do not break bit-flip protection. 
We measure the scaling of errors for $|\alpha|^2\in[1.4, 11.3]$, a range on which we can measure both the phase-flip and bit-flip rates (Fig.~4b,c). We observe the bit-flip time multiply by 4.2 for every added photon, culminating at 15 seconds. Importantly, in the presence of the continuous memory drive, the bit-flip only slightly reduces, remaining above 10 seconds for $|\alpha|^2=11.3$. On the other hand, the measured dephasing rate ${T_Z}^{-1}$ increases linearly with $|\alpha|^2$, closely following the theoretical prediction $\Gamma^{\kappa_a}_Z=2\kappa_a |\alpha|^2$. Notably, the oscillator lifetime extracted from a linear fit to the data is 25$\%$ larger than the one obtained from spectroscopy, possibly due to the interplay of the strong two-photon pump and uncontrolled parametric processes. 

{In conclusion, this experiment demonstrates quantum tomography and coherent control of a cat-qubit without breaking bit-flip protection up to bit-flip times of 10 seconds. This constitutes a $10^4$ improvement over other cat-qubit implementations, and a $10^6$ enhancement over the oscillator lifetime. We measure a phase-flip time of 490~ns, and perform a $\pi$ rotation around the $Z$ axis in 235~ns. Although we achieved $g_2/\kappa_a\approx 80 \gg 1$, this ratio needs to be further increased to improve measurement fidelities and reduce state preparation and gate errors below the error correction threshold \cite{Guillaud2019, Gautier2022}. Possible directions for progress are circuit engineering to increase $g_2$ \cite{Aiello2022, Marquet2023}, optimized gate design \cite{Eickbusch2022,Gautier2023} and the integration of recent advances in nanofabrication \cite{Place2021, Wang2022, Kono2023} in order to improve our oscillator lifetime by at least one order of magnitude. With these improvements in hand, we envision assembling multiple cat-qubits in hardware-efficient error correcting architectures \cite{Guillaud2019, Puri2020, Darmawan2021, Chamberland2022}, and operating them to correct phase-flips without breaking bit-flip protection.}

\paragraph*{\bf Acknowledgements }
We thank N. Frattini for his insight on applying the protocol of Albert {\it et. al} Phys. Rev. Lett. (2016) to Wigner tomography. We thank the SPEC at CEA Saclay for providing nano-fabrication facilities. This work was supported by the QuantERA grant QuCOS, by ANR 19-QUAN-0006-04. This project has received funding from the European Research Council (ERC) under the European Union’s Horizon 2020 research and innovation programme (grant agreements No.\ 851740 and No. 884762). This work has been funded by the French grants ANR-22-PETQ-0003 and ANR-22-PETQ-0006 under the 'France 2030 plan'. This research is partially funded by the CATQUBIT Horizon Europe project (grant agreement 190110172).

\paragraph*{\bf Data availability }
The data that support the findings of this work are available from the corresponding author upon a request.

\paragraph*{\bf Code availability }
The code used for data acquisition, analysis and visualization is available from the corresponding author upon a request.

\paragraph*{\bf Author contributions}

P.C-I, R.L, S.J and Z.L conceived the experiment. U.R and A.B designed the chip with M.H and F.R providing support. U.R and A.B. measured the device and analyzed the data. E.A and N.P fabricated the chip. R.G, J.C, A.M, L.-A.S, P.R, A.S and M.M provided theory support. U.R, A.B and Z.L wrote the paper with input from all authors.

\paragraph*{\bf Corresponding author}
Correspondence should be addressed to Z. Leghtas: zaki.leghtas@ens.fr

\paragraph*{\bf Ethics declarations - Competing interests}
Authors affiliated with Alice \& Bob (A\&B) have financial interest in the company. ZL, MM and PCI are shareholders of A\&B. PCI is a part time employee of A\&B.

\bibliography{biblio}


\textbf{\Large Supplementary Material}
\setcounter{equation}{0}
\setcounter{section}{0}
\setcounter{figure}{0}
\setcounter{table}{0}
\setcounter{page}{1}

\renewcommand{\theequation}{S\arabic{equation}}
\renewcommand{\thefigure}{S\arabic{figure}}
\renewcommand{\thesection}{S\arabic{section}}
\renewcommand{\thetable}{S\arabic{table}}

\section{Device fabrication and wiring}
\begin{figure}
\centering
\includegraphics[width=1\columnwidth]{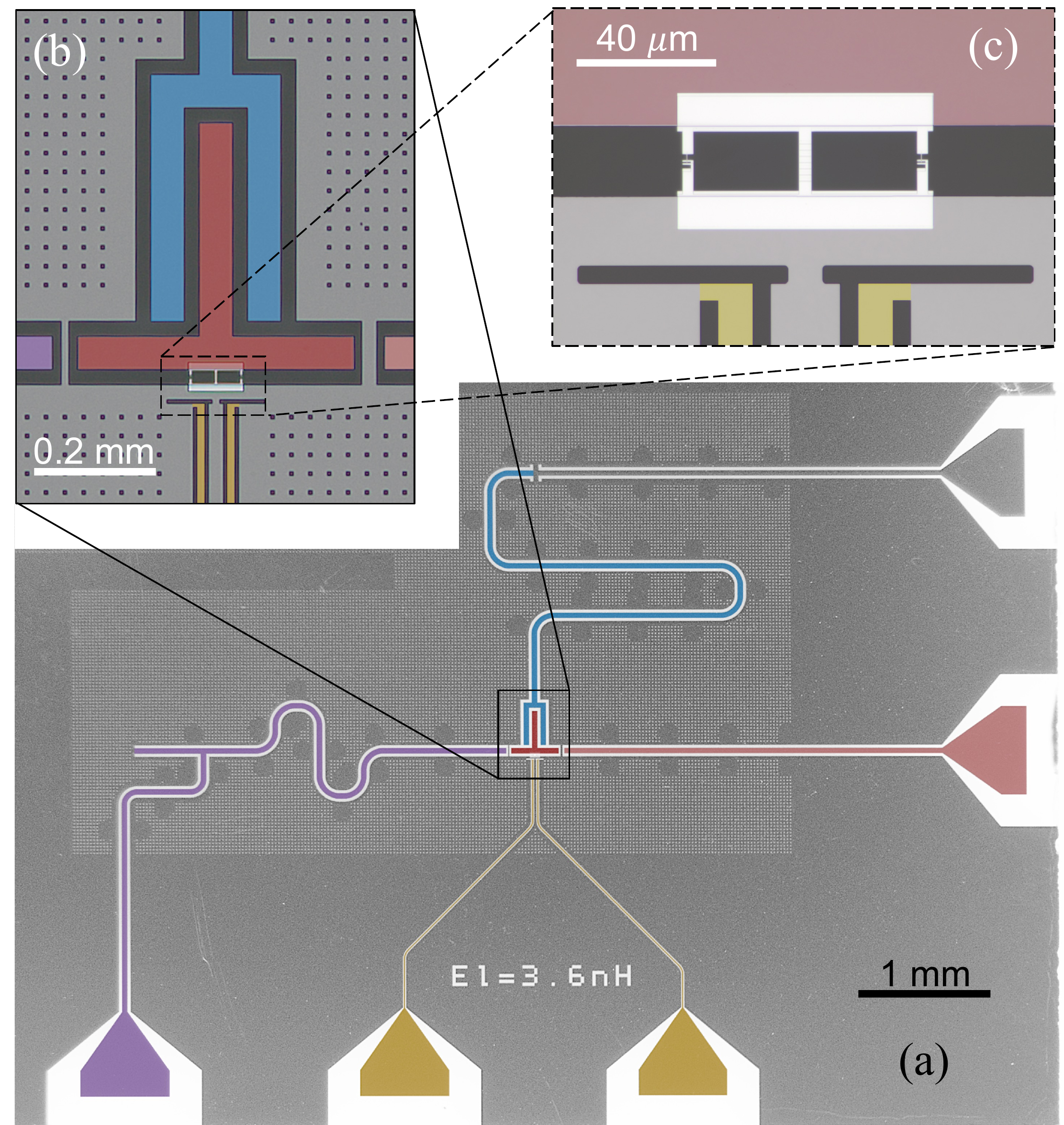}
\caption{{\bf Optical micrographs of a twin sample.} (a) A quarter wave-length coplanar waveguide (CPW) resonator supports the memory mode (blue). It is shunted to ground on its top end. The transmission line in its vicinity is unused in our experiment and is shunted to ground with a wirebond. On its bottom end, the memory oscillator couples capacitively to a superconducting island (red), corresponding to the buffer capacitance. The buffer is coupled from the left to a transmission line (purple) incorporating a Purcell filter. From this port we collect the buffer signals, and inject the buffer drive. A weakly coupled transmission line (pink) channels the memory drive and the cancellation tone for the longitudinal coupling into the circuit. The two symmetric lines (yellow) carry common and differential flux to bias the ATS, as well as the two-photon and longitudinal pump signals. (b) Zoom on the buffer island (red) and its surroundings. The tantalum film (light grey) is punctured with holes to avoid vortex trapping. The substrate (dark grey) is made out of sapphire. (c) Zoom on the ATS, that is fabricated in aluminum (white). The two small cross junctions (left and right) form a SQUID loop that is shunted in its center by a chain of 13 junctions.}
\label{fig:supmat_chip}
\end{figure}

\begin{figure*}
\centering
\includegraphics[width=1.5\columnwidth]{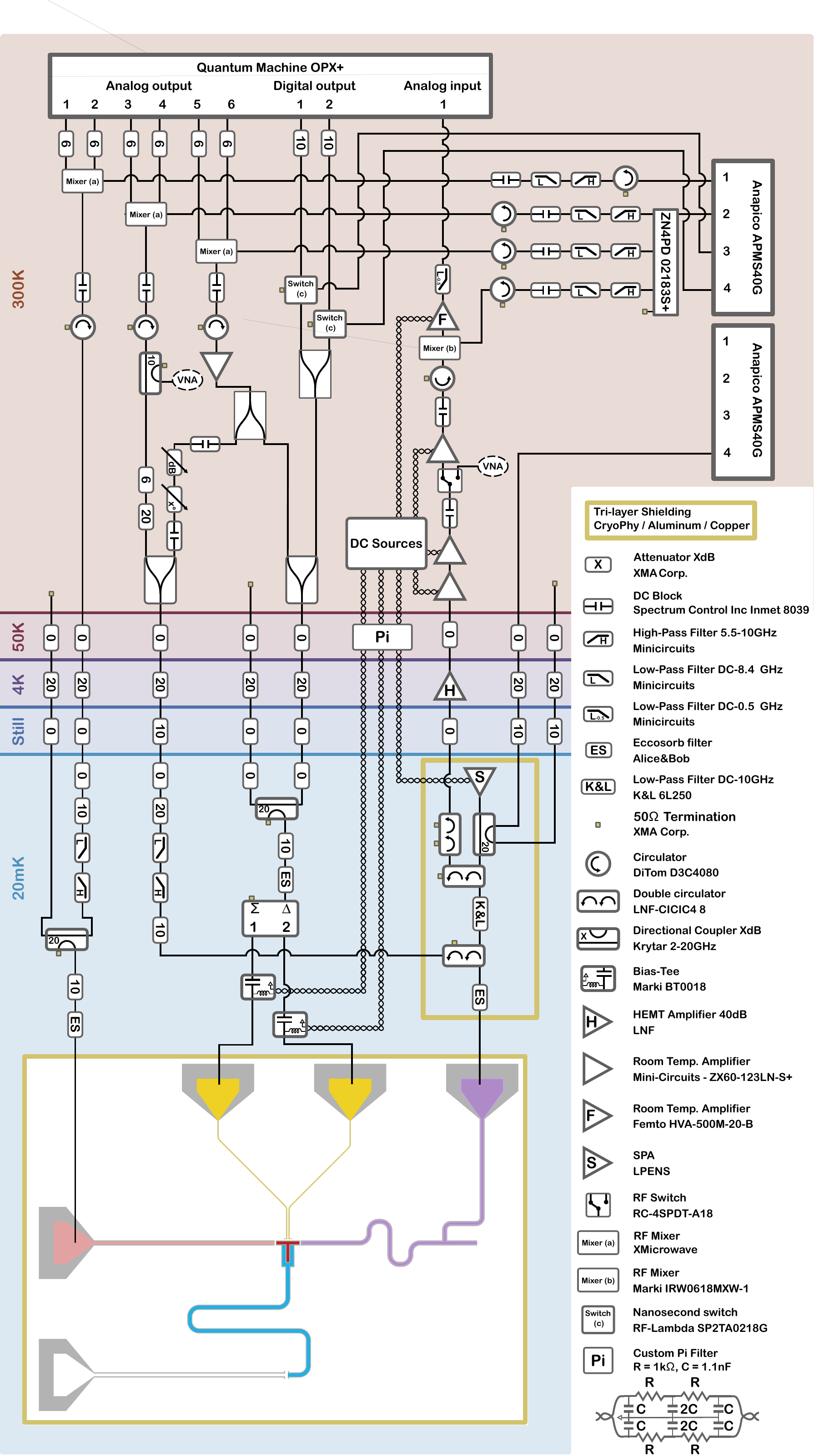}
\caption{Wiring diagram of the experiment.}
\label{fig:supmat_frigo}
\end{figure*}

In this section we describe the recipe we follow to fabricate the sample displayed in Fig.~\ref{fig:supmat_chip}. The wiring diagram of the experiment is displayed in Fig.~\ref{fig:supmat_frigo}.

\paragraph{Wafer preparation} 
Our wafer is purchased from Starcryo and consists of a 2-inch, 430~$\mu$m-thick sapphire wafer, on which 200~nm of $\alpha$-phase tantalum were sputtered.

\paragraph{Circuit patterning} 
Large features of the circuit ($>5$ µm) are patterned using direct laser lithography, after coating with a 500~nm layer of positive resist (S1805). The exposed sample is developed in MF319 (1 min) and rinsed in deionized water (1 min), after which reactive ion etching (SF6, 0.02~mbar, 100 W) is used to transfer the pattern onto the Tantalum layer. Finally, the resist is stripped in an acetone bath (50°C, 20 min + 10 min sonication), followed by a 3 min oxygen plasma (0.02 mbar, 200 W).

\section{Circuit Hamiltonian}

\paragraph{Junction patterning}
On top of the Tantalum circuit, Josephson junctions are patterned using the Dolan bridge technique. 
We coat the wafer with a bilayer consisting of MAA EL13 (650 nm, baked 3 min at 185°C) and PMMA A3 (120 nm, baked 30 min at 185°C). On top, a thin layer (20 nm) of conductive resist (Electra92) is spinned to improve charge-evacuation during lithography. Josephson junctions are patterned via electron beam lithography (20 keV). The wafer is developed in a 3:1, IPA:H$_2$O bath (6 °C, 2 min), and then rinsed in IPA (10 s) and blow-dried. Finally, a step of oxygen plasma ashing (10 s, 200 mbar, 75 W) allows to clean the mask from residual organic contaminants.

\paragraph{Junction deposition}
Josephson junctions are evaporated via electron beam double-angle evaporation. First, a step of Argon milling (15 s, 5 mA, 500 V, $\pm$ 30°) is used to remove native Tantalum oxyde and assure good electrical contact between junctions and the Tantalum circuit. Then, two layers of aluminium (35 nm, 70 nm) are evaporated at angles $\pm$ 30°, with a step of static oxidation in between (15 O$_2$ - 75 Ar, 20 mbar, 10 min). Lift-off of the excess metal is done in acetone (50°C, 1 h, 2 min sonication) and an NMP-based remover (50°C, 10 min).

\section{Circuit Hamiltonian}

Our circuit, displayed in Fig.~\ref{fig:supmat_chip}, is well described by the following Hamiltonian \cite{Lescanne2020}:
\begin{equation}
\label{eq:Htot}
\begin{split}
    \bH_\text{circuit} &= \hbar\omega_{a,0} \ba^\dag \ba+ \hbar\omega_{b, 0} \bb^\dag \bb \\
    &-2E_J \cos(\varphi_\Sigma)\cos(\bp + \varphi_\Delta)\\
    &+2\Delta E_J \sin(\varphi_\Sigma)\sin(\bp + \varphi_\Delta)\\
    \bp&=\varphi_a(\ba+\ba^\dag)+\varphi_b(\bb+\bb^\dag)\;.
\end{split}
\end{equation}
The ATS is composed of two Josephson junctions of Josephson energies $E_{J1}, E_{J2}$ in a SQUID loop configuration. We define $E_J=\frac{1}{2}(E_{J1}+E_{J2})$ and $\Delta E_J=\frac{1}{2}(E_{J1}-E_{J2})$. An inductive shunt splits the SQUID in a left and right loop that we thread with magnetic flux $\varphi_{L,R}$. We define the common and differential flux as $\varphi_{\Sigma,\Delta}=\frac{1}{2}(\varphi_{L}\pm\varphi_{R})$. Due to inevitable Josephson junction asymmetry $\Delta E_J$, the memory and buffer angular frequencies $\omega_{a,0},\omega_{b,0}$ slightly deviate from the mode frequencies $\omega_{a,\pm},\omega_{b,\pm}$ at the saddle points $(\varphi_\Sigma,\varphi_\Delta)=(\pi/2,\pm\pi/2)$. Indeed, $\omega_{a/b,\pm}\approx\omega_{a/b,0}\mp2\Delta E_J\varphi_{a/b}^2$, where $\varphi_{a,b}$ are the zero point phase fluctuations of the buffer and memory across the ATS.

\begin{figure}
\centering
\includegraphics[width=1\columnwidth]{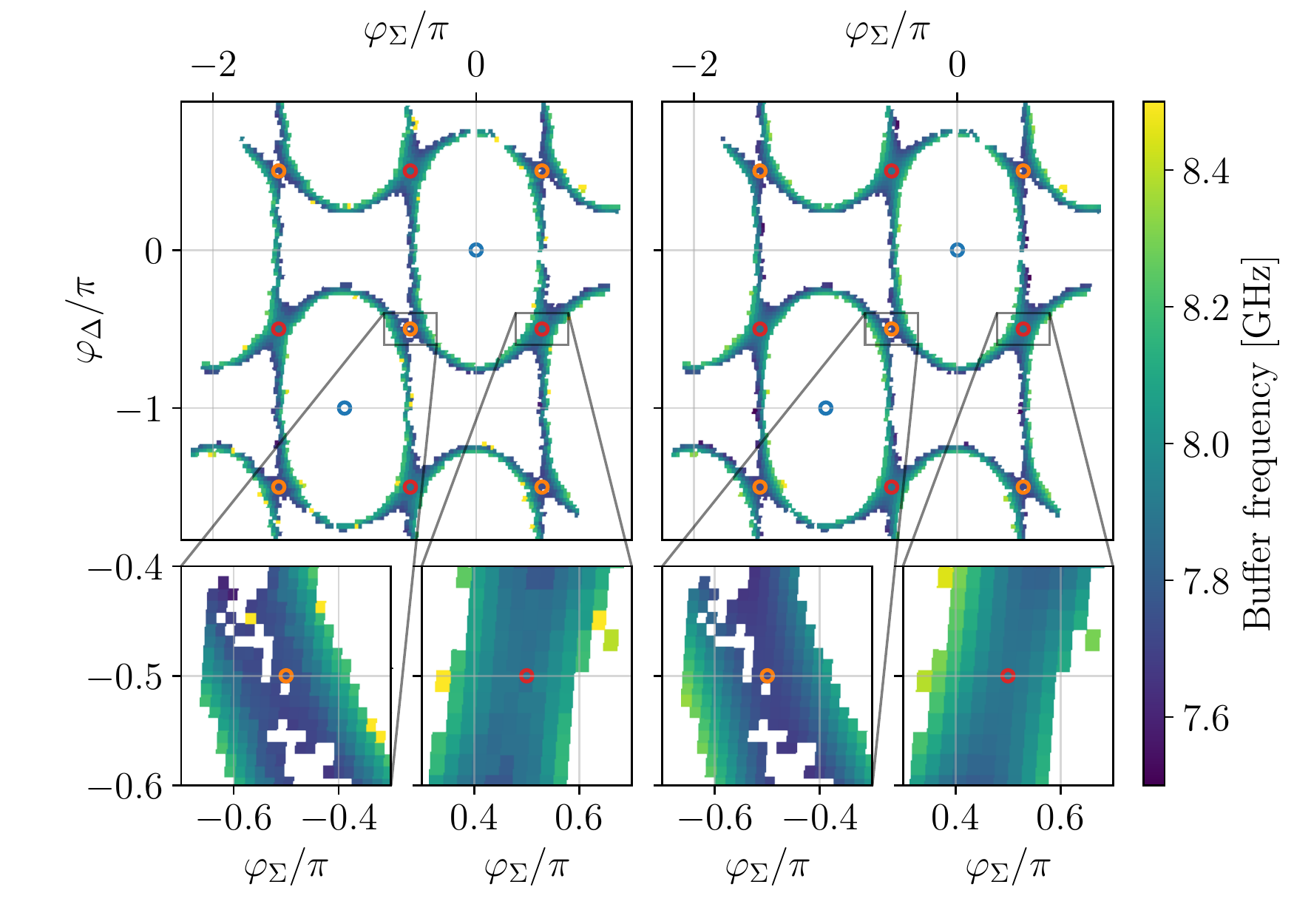}
\caption{{\bf Flux maps.} Measured (left) and simulated (right) buffer frequency (color) as a function of differential (x-axis) and common (y-axis) flux. Open orange and red circles mark the location of the two non-equivalent saddle points. The blue open circles point to the maximum of the flux map. The bottom panels are zooms around two nonequivalent saddle points. In the blank regions, the buffer is misaligned with its Purcell filter and is therefore too undercoupled to the transmission line to be detected. For the rest of the experiment, the flux is set to  $(\varphi_\Sigma,\varphi_\Delta)=(-\pi/2,-\pi/2)$.}
\label{fig:supmat_fluxmap}
\end{figure}

In order to pin down the value of each parameter entering this Hamiltonian, we run a 3D finite element electromagnatic simulation of our device in the absence of the two SQUID junctions, and where the ATS inductance consisting in a chain of 13 junctions is modeled by an inductive lumped element of energy $E_L$. Although room temperature resistance measurements provide an estimate of $E_L$ through the Ambegaokar-Baratoff formula, its precise value is determined by sweeping $E_L$ in the simulation and extracting the corresponding $\omega_{a,0}, \omega_{b,0}, \varphi_a, \varphi_b$. We then pick the value of $E_L$ for which $\omega_{b,0}$ matches best the mean of the measured frequencies $\omega_{b,\pm}$ at the two saddle points. The difference between the measured frequencies at these two saddle points then sets the SQUID junctions asymmetry $\Delta E_J$. Finally fitting the measured buffer frequencies away from the saddle points sets the value of $E_J$. The comparison between the measured buffer frequencies and the ones extracted from Eq.~\eqref{eq:Htot} is displayed in Fig.~\ref{fig:supmat_fluxmap}. The parameters entering the Hamiltonian are listed in Table~\ref{table:supmat_params}.

\begin{center}
\begin{table}
\begin{tabular}{ |c|c| } 

\hline
$\omega_{b,0}/2\pi$ & 7.70 GHz\\ 
\hline
$\varphi_{b}$ & 0.29 \\
\hline
    \hline
$\omega_{a,0}/2\pi$ & 5.26 GHz\\ 
\hline
$\varphi_{a}$ & 0.06 \\
\hline
\hline
$E_L/h$ & 42.76 GHz \\ 
\hline
$E_J/h$ & 12.03 GHz \\ 
\hline
$\Delta E_J/h$ & 0.47 GHz \\ 
\hline
\end{tabular}
\caption{Buffer, memory and ATS parameters entering Hamiltonian \eqref{eq:Htot} in order to produce the fit of Fig.~\ref{fig:supmat_fluxmap}}
\label{table:supmat_params}

\end{table}
\end{center}

\section{Calibration of parametric pumps}

\subsection{Parametric pumping of the ATS}
\label{sec:parametric_pumping_ATS}
In this experiment, the ATS is employed to mediate both a longitudinal coupling between the memory and the buffer, and to induce two-photon dissipation in the memory. To this end, we navigate the map displayed in Fig.~\ref{fig:supmat_fluxmap} towards the saddle point $(\varphi_\Sigma,\varphi_\Delta)=(-\pi/2,-\pi/2)$, and keep this value fixed for the remainder of the experiment. At this bias point, the memory and buffer are first-order flux-insensitive and inherit small spurious Kerr interactions. For simplicity, we set $\Delta E_J$ to zero in this analysis and hence $\omega_{a/b,0}=\omega_{a/b,\pm}$. In the following, we denote $\omega_{a/b}$ the memory and buffer angular frequencies.

We combine this DC flux bias with an AC common flux pump of amplitude $\varepsilon_p$ and frequency $\omega_p$, yielding $\varphi_\Sigma=-\pi/2+\varepsilon_p\cos(\omega_p t)$ and $\varphi_\Delta=-\pi/2$.  Injecting these expressions in Eq.~\eqref{eq:Htot}, we recover a nonlinear interaction term between the memory and buffer of the form $U(\bp)=-2E_J\sin(\varepsilon_p\cos(\omega_p t))\sin(\bp)$. In the regime where the pump amplitude $\varepsilon_p\ll 1$, $U(\bp)\approx-2E_J\varepsilon_p\cos(\omega_p t)\sin(\bp)$. Moreover since $\varphi_{a,b}\ll 1$, we may expand $\sin(\bp)$ to third order : 

\begin{equation}
\label{eq:supmat_U}
U(\bp)\approx-2E_J\varepsilon_p\cos(\omega_p t)(\bp-\frac{1}{6}\bp^3)\;.
\end{equation}

\subsection{Longitudinal coupling}
\label{sec:longitudinal_coupling}
In this section we describe how the ATS is pumped to mediate a longitudinal coupling between the memory and the buffer, a crucial ingredient for our transmon-free quantum tomography protocol.

The amplitude of the pump tone employed for longitudinal coupling is denoted $\varepsilon^l_p$, and its frequency is to $\omega_p^l\approx\omega_b$. Performing a rotating wave approximation, the coupling term of Eq.~\eqref{eq:supmat_U} reduces to 
\begin{eqnarray}
    U_{\omega_p^l=\omega_b}(\bp)/\hbar&\approx&g_\text{lin}(\bb + \bb^\dagger) + g_l \ba^\dagger \ba (\bb + \bb^\dagger)\notag\\
    &+& g_\text{sp}\left( {\bb^\dagger}^2 \bb+\bb^\dagger\bb^2\right)\;.
\end{eqnarray} 
The first term is a linear drive on the buffer of amplitude $g_\text{lin}=-E_J\varepsilon^l_{p}\varphi_b/\hbar$ (third order corrections in $\varphi_{a,b}$ are neglected). We cancel this term by applying an additional tone on the buffer, with equal amplitude and opposite phase, through a capacitively coupled port. The second term is the desired longitudinal coupling of strength $g_l = E_J \varepsilon^l_p \varphi_a^2 \varphi_b/\hbar$. Finally, the third term, where $g_\text{sp}=\frac{1}{2}\varphi_b^2/\varphi_a^2 g_l$, is a spurious term that limits the dynamical range of our detector. Future experiments will optimize the ratio $\varphi_b/\varphi_a$ to maximize the readout fidelity of the memory through this longitudinal coupling to the buffer. Considering only the longitudinal interaction, when there are exactly $n_a$ photons in the memory, the buffer converges to a coherent state of amplitude $\beta_{1ph}\times n_a$, where $\beta_{1ph}=-2ig_l/\kappa_b$.

\begin{figure}
\centering
\includegraphics[width=1\columnwidth]{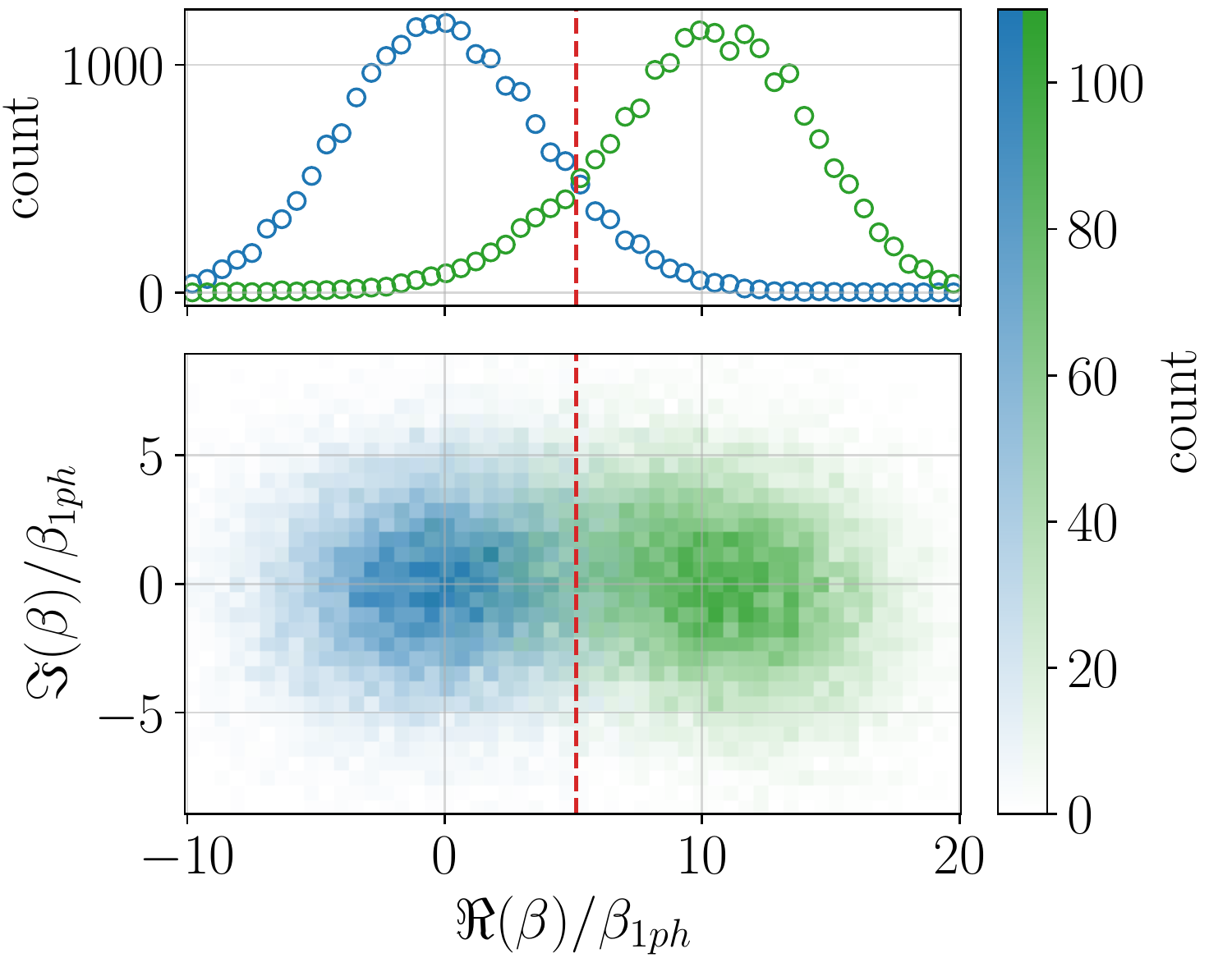}
\caption{{\bf Measurement fidelity of the longitudinal readout.} The memory is prepared in the vacuum (blue) or displaced to a coherent state containing an average photon number of 10 (green). The longitudinal pump is then activated, and the buffer output field is integrated over 10~$\mu$s. This measurement is repeated 20 000 times for each memory state preparation. (Bottom) Two-dimensional histogram of the real and imaginary quadratures of the buffer field. The histogram was rotated so that all the information on the memory state is encoded in the real quadrature. The histogram was also displaced so that the origin of the axis coincides with the center of the histogram corresponding to a memory in the vacuum. Finally the real and imaginary axis are re-scaled in units of average photon number present in the memory. (Top) Histogram of the real quadrature of the buffer output field, obtained by integrating the two-dimensional histogram over the imaginary quadrature. The vertical red dashed line is used to compute a single shot fidelity. We denote $P(L|0ph)$ the probability of measuring a data point on the left of the threshold, given a memory state prepared in the vacuum. Similarly, we denote $P(R|10ph)$ the probability of measuring a data point on the right of the threshold, given a memory state prepared in a coherent state containing 10 photons. We define the measurement fidelity as $(P(L|0ph)+P(R|10ph))/2$, and find 89\%.
}
\label{fig:supmat_fidelity}
\end{figure}

In the following we describe the procedure we follow to calibrate (i) the pump frequency (ii) the amplitude of the cancellation tone and (iii) its phase. We sweep these parameters in a three-dimensional grid. For each combination we activate the pump and cancellation tones and record histograms of the buffer output both with and without an initial memory displacement. We select the combination of parameters that maximizes the Kullback–Leibler divergence between the two histograms. Finally, we record the buffer output as a function of the amplitude of the memory displacement (Fig.~1.d). We observe that the buffer output amplitude grows linearly with the number of photons in the memory up to about 5 photons, as expected from a longitudinal interaction. Above 5 photons, we observe a deviation from the linear dependence likely due to the spurious term $g_\text{sp}\left( {\bb^\dagger}^2 \bb+\bb^\dagger\bb^2\right)$. The power of the pump is chosen to maximize the span covered by the buffer amplitude over a window of 0 to 10 memory photons, which is the range of interest in this experiment. For the optimal pump power, we are able to distinguish between the vacuum and a coherent state in the memory containing 10 photons with a single-shot fidelity of 89\% (Fig.~\ref{fig:supmat_fidelity}).

\subsection{Two-photon dissipation}

\begin{figure}
\centering
\includegraphics[width=1\columnwidth]{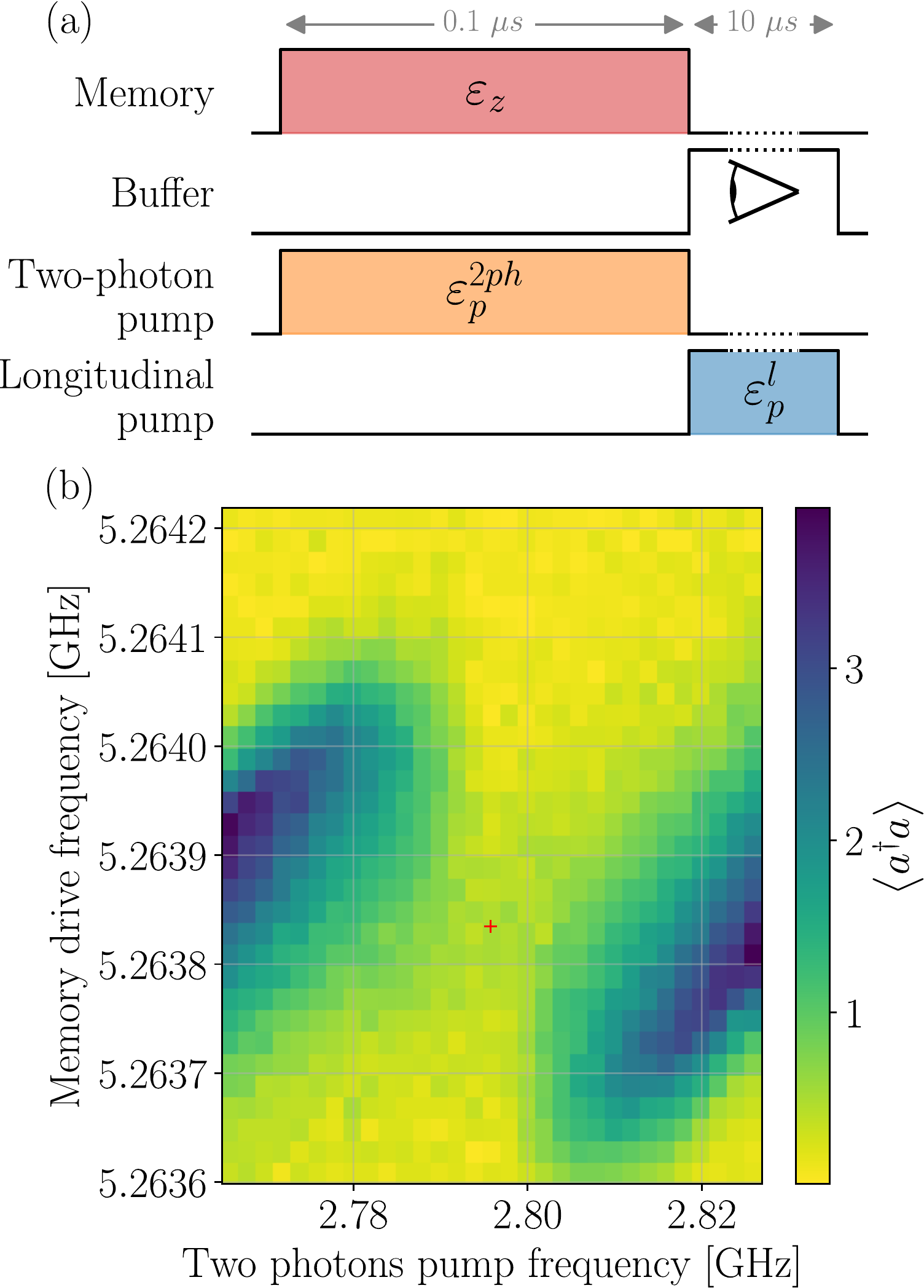}
\caption{{\bf Calibration of the two-photon pump and memory drive frequencies.} (a) Pulse sequence: square pulses are applied on both the memory and the two photon pump. Their duration is $0.1~\mu\text{s}$ and their frequencies are swept. Then we measure the memory photon number through longitudinal coupling for $10~\mu\text{s}$ (b) Photon number in the memory (color) as a function of the two-photon pump frequency  $\omega_{p}^{2ph}$ (x-axis) and memory drive frequency $\omega_Z$ (y-axis). At the saddle point (red cross), the frequency matching condition $2\omega_Z-\omega_p^{2ph}=\omega_b$ is satisfied. Subsequently we set the buffer drive frequency to $\omega_d=2\omega_Z-\omega_p^{2ph}$.}
\label{fig:supmat_reversed_diamond}
\end{figure}

In this section we describe how we implement two-photon dissipation in the memory, the mechanism at the core of the cat-qubit. The amplitude of the pump tone employed for two-photon dissipation is denoted $\varepsilon_p^{2ph}$ and its frequency is set to $\omega_p^{2ph}=2\omega_a-\omega_b$. Performing a rotating wave approximation on Eq.~(1), we find
\begin{equation*}
U_{\omega_p^{2ph}=2\omega_a-\omega_b}(\bp)/\hbar=g_2 \left({\ba^2\bb}^\dag+{\ba^\dag}^2\bb\right)\;.
\end{equation*}
We recover the two-to-one exchange mechanism described in Eq.~(1), where $g_2=\frac{1}{2}E_J\epsilon_p^{2ph}\varphi_a^2\varphi_b/\hbar$. For simplicity, $g_2$ is assumed to be a real number. Note that $g_l\propto g_2$, where $g_l$ is the longitudinal coupling derived in the previous section, and hence the performances of the longitudinal readout and the cat-qubit stabilization are directly linked. In the following, we simulate the dynamics of our two-photon dissipative system by numerically solving the Lindblad master equation: 

\begin{eqnarray}
\label{eq:master_equation}
\frac{d}{dt}{\rho} &=& -i\left[g_2\left( \ba^2 \bb^\dagger+{\ba^\dagger}^2\bb \right)-\varepsilon_d^*\bb-\varepsilon_d\bb^\dag, \rho\right]\\ 
&+& \kappa_a(1+n^{th}_a)D[\ba]\rho + \kappa_a n^{th}_aD[\ba^\dagger]\notag\\
&+& \kappa_b(1+n^{th}_b)D[\bb]\rho + \kappa_b n^{th}_b D[\bb^\dagger]\notag\;,
\end{eqnarray}

where, $\rho$ is the buffer-memory density matrix, and $n^{th}_{a,b}$ are the memory and buffer thermal occupation numbers. 

\subsubsection{Frequency calibration}

{The two-photon pump Stark-shifts the mode frequencies $\omega_{a,b}$. To calibrate the shifted frequencies}, we activate the pump and vary its frequency $\omega_p^{2ph}$ in the vicinity of the frequency matching condition $2\omega_a-\omega_b$, while applying a weak drive on the memory (which will later be used to perform Z rotations) of amplitude $\epsilon_Z$ whose frequency $\omega_Z$ is varied around $\omega_a$. For each value of this two-dimensional frequency sweep, we measure the memory photon number through longitudinal coupling to the buffer. This defines a two-dimensional map shown in Fig.~\ref{fig:supmat_reversed_diamond}, where $(\omega_p^{2ph},\omega_Z)=(2\omega_a-\omega_b,\omega_a)$ is a saddle point. Indeed, at this operating point, the memory photon number is a balance between the memory drive strength, and the two-photon confinement rate. Detuning the two-photon pump results in a lower confinement rate, and hence a larger memory photon number. Conversely, detuning the memory drive reduces the memory photon number. Fitting the data to a second degree polynomial, we precisely locate the saddle point and pin down $\omega_p^{2ph}$ and $\omega_Z$. Finally, we set the buffer drive of amplitude $\epsilon_d$ to frequency $\omega_d=2\omega_Z-\omega_p^{2ph}$.

\begin{center}
\begin{table}
\begin{tabular}{ |c|c| } 
\hline
$g_2/2\pi$ & 0.763 MHz\\ 
\hline
$\kappa_b/2\pi$ & 2.6 MHz \\
\hline
$\kappa_a/2\pi$ & 9.3 kHz\\ 
\hline
$n_{a}^{th}$ & 10\% \\
\hline
$n_{b}^{th}$ & 1.1\% \\ 
\hline
\end{tabular}
\caption{Parameters entering a Lindblad master equation to produce the numerical simulations of the paper, namely Fig.~1c, Fig.4b,  Fig.~\ref{fig:supmat_deflation}b, Fig.~\ref{fig:supmat_rabi_zeno}c and Fig.~\ref{fig:supmat_states_preparation}b.}
\label{table:supmat_qutip_params}

\end{table}
\end{center}

\subsubsection{Two-photon exchange rate calibration}

\begin{figure}
\centering
\includegraphics[width=1\columnwidth]{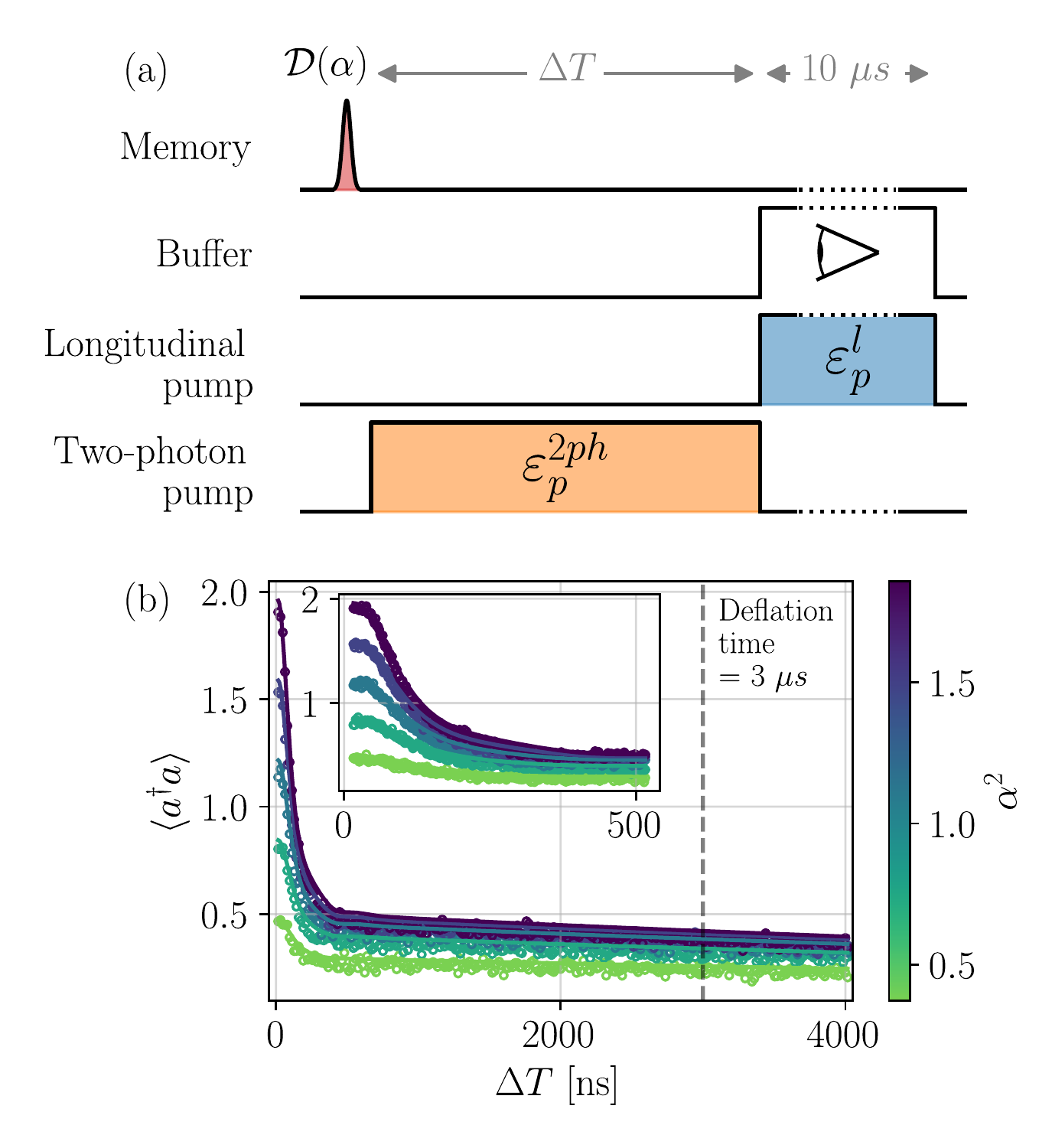}
\caption{{\bf Deflation time.} Calibration of the convergence time towards the $\ket{\pm}_{\alpha=0}$ manifold, colloquially referred to as the deflation time. (a) Pulse sequence : first, a coherent state of amplitude $\alpha$ in the memory is created. A two-photon pump is then activated for variable time $\Delta T$, and finally, the memory photon number is measured by longitudinal readout through the buffer. (b) Photon number (y-axis) as a function of $\Delta T$ (x-axis) for various initial coherent states $\ket{\alpha}$ (color). The data (open circles) are overlaid on theory curves obtained from solving the Lindblad master equation of  Eq.~\eqref{eq:master_equation} with $\varepsilon_d=0$. The parameters entering the simulation are displayed in Table~\ref{table:supmat_qutip_params}. The only parameters that were varied to fit the data are  $g_2$ and $n_b^{th}$. Other parameters were determined independently.}
\label{fig:supmat_deflation}
\end{figure}

{To determine $g_2$, we measure the rate at which photons are removed from the memory mode by the two-photon dissipation mechanism. To do so, we} implement the following pulse sequence (Fig.~\ref{fig:supmat_deflation}a): first the memory is displaced to a coherent state of amplitude $\alpha$, then the two-photon pump is activated for a variable time $\Delta T$, and finally, the memory photon number is measured by activating the longitudinal pump. The measured memory photon number versus time for various $\alpha$ is displayed in Fig.~\ref{fig:supmat_deflation}b. Two timescales are apparent. On a short timescale $1/\kappa_\text{conf}\approx2/\kappa_b=120$~ns, pairs of photons are quickly removed from the memory until it is left in a mixture of $|0\rangle$ or $|1\rangle$ Fock states. At this stage, the number of photons $\langle a^\dag a\rangle$ ranges from $1/2$ for initial state amplitudes $\alpha\gg1$ down to $0$ when $\alpha=0$. Then, the remaining photon is dissipated into the memory bath on a longer timescale $1/\kappa_a$. The data are fitted by solving the master equation of the memory-buffer system (Eq.~\eqref{eq:master_equation}), with $\epsilon_d=0$, and $\rho(0)=\ket{\alpha}\bra{\alpha}$. The only fit parameters are the two-photon exchange rate and the thermal population of the buffer, for which we obtain $g_2/2\pi=763$~kHz and $n^{th}_b=1.1\%$. The dissipation rates $\kappa_a$ and $\kappa_b$ are measured independently using standard techniques. The number of photons in the memory (y-axis) and the thermal population of the memory $n^{th}_a$ are calibrated through Wigner tomography of the memory (Sec.~\ref{sec:photon_number_calibration}).

\subsubsection{Cat state preparation}

\begin{figure}
\centering
\includegraphics[width=1\columnwidth]{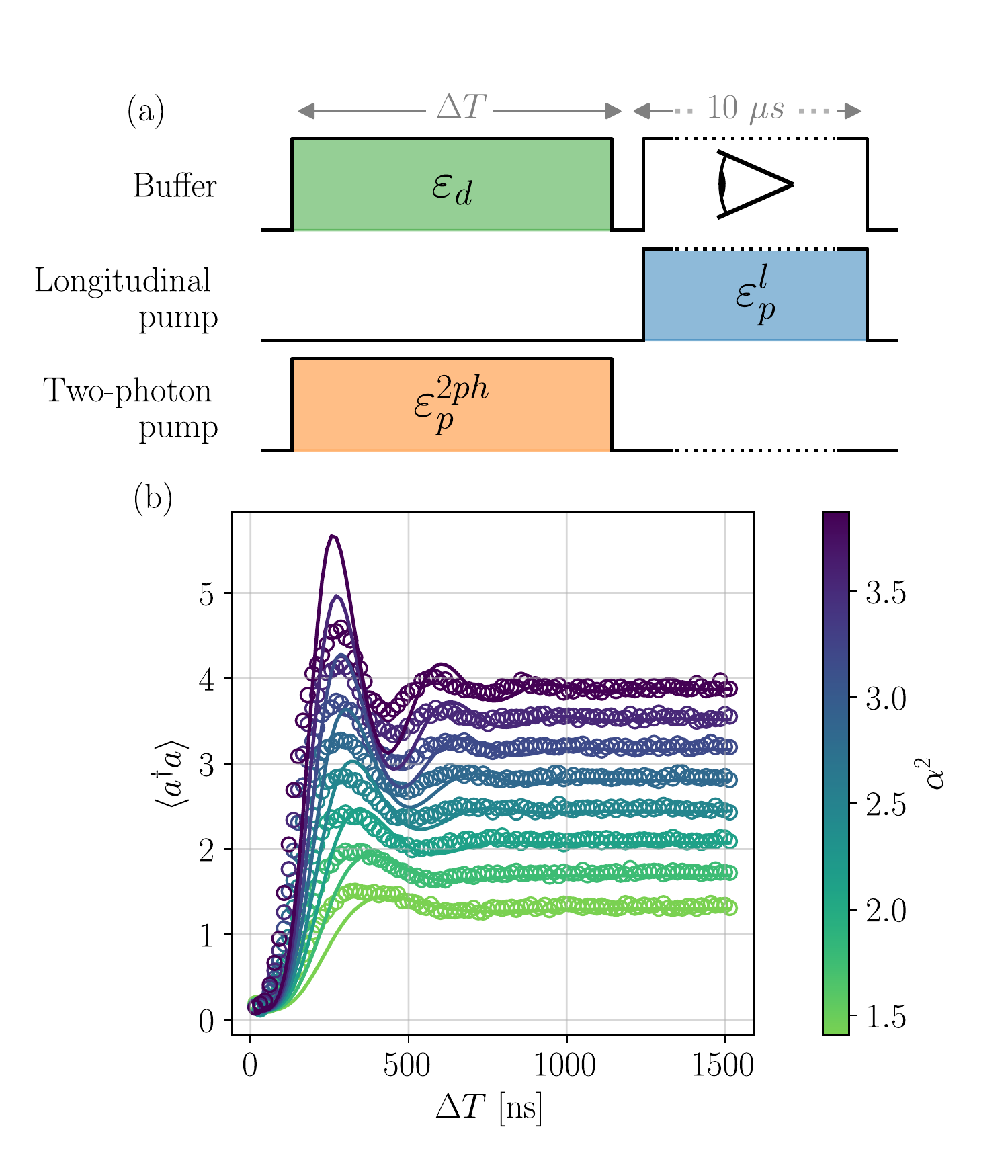}
\caption{{\bf Inflation time.} Calibration of the convergence time towards the $\ket{\pm}_{\alpha}$ manifold, colloquially referred to as the inflation time. (a) Pulse sequence : starting from the vacuum in the memory, the two-photon pump and buffer drive is activated for variable time $\Delta T$. The memory photon number is then measured by longitudinal readout through the buffer. (b) Photon number (y-axis) as a function of $\Delta T$ (x-axis) for drive amplitudes targeting various steady state memory photon numbers (color).  The data (open circles) are overlaid with the results of a numerical simulation of Eq.~\eqref{eq:master_equation} with $\rho(0)=\ket{0}\bra{0}$,  $\epsilon_d=\alpha^2 g_2$ and no fit parameters. The parameters entering this simulation are summarized in Table.~\ref{table:supmat_qutip_params} have all been independently calibrated.}
\label{fig:supmat_inflation}
\end{figure}

Since in this experiment we have entirely removed the
transmon, we rely solely on the two-to-one photon exchange mechanism to generate quantum states. To this
end, starting from the vacuum in the memory, we activate
for a varying time $\Delta T$ both the two-photon pump and
a buffer drive at various amplitudes $\varepsilon_d$. We then measure
the memory photon number through longitudinal
coupling to the buffer. The data are reported in Fig.~\ref{fig:supmat_inflation}.
Three timescales are apparent. For $\Delta T \in [0, 
250~\text{ns}]$, we observe a rapid increase of photon number in the memory. Indeed, single photons from the buffer drive
combine with the pump and pairs of photons are squeezed
into the memory. For $\Delta T \in [250~\text{ns}, 600~\text{ns}]$, we observe
damped oscillations in the memory photon number. Although
we do not quantitatively understand the amplitude of
these oscillations, they can be reproduced in simulation
by adding a Kerr on the buffer, and a detuning of the
pump and drive from the frequency matching condition.
For $\Delta T \ge
600~\text{ns}$, the photon number reaches a plateau.
This is expected when the memory has converged into the
quantum manifold spanned by $\ket{\pm}_\alpha$. The specific state
that is reached within this manifold depends on the initial
state. Since the memory is initialized in the vacuum,
the memory converges to the even cat $\ket{+}_\alpha$. However,
due to inevitable losses in the memory, this state will
evolve into a classical statistical mixture of $\ket{\pm}_\alpha$ at a
rate $2|\alpha|^2\kappa_a$. The fidelity of preparation of $\ket{+}_\alpha$ is therefore
maximized for the smallest time $\Delta T$ that insures the
convergence of the memory in its steady state manifold. The cat states generated for Wigner tomography in Fig.~2c and Fig.~3a,b were prepared in $\Delta T=1~\mu$s. Those for the measurements of $\langle \bX_\alpha \rangle$ in Fig.~3c and Fig.~4a were prepared in $\Delta T=400$~ns.

\subsection{Active memory reset}
\label{sec:reset}

\begin{figure}
\centering
\includegraphics[width=0.8\columnwidth]{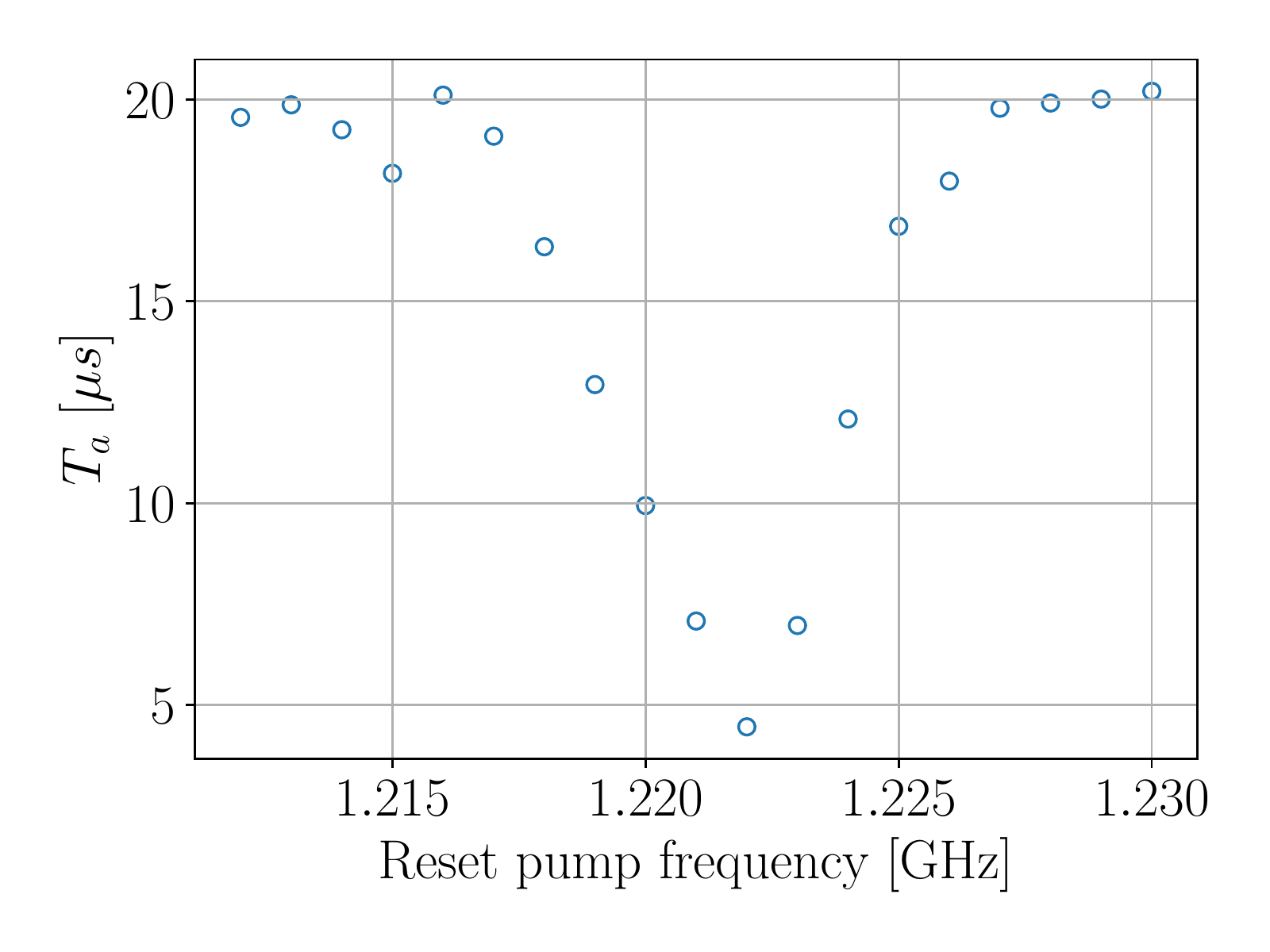}
\caption{{\bf Calibration of the reset pump frequency.} Memory lifetime $T_a$ (y-axis) versus the reset pump frequency (x-axis). The data (open circles) are extracted from a fit of the form $\exp(-t/T_a)$ of the memory photon number decay versus time $t$ in the presence of the reset pump.}
\label{fig:supmat_active_reset}
\end{figure}

In the main text we acquire highly averaged and high resolution Wigner functions. The protocol for the generation and Wigner tomography of a memory state consists in a $\approx 15 \mu$s pulse sequence followed by a long wait time for the memory to relax back to the vacuum. This wait time is typically $\approx 5/ T_a$, where $T_a=1/\kappa_a$ is the memory lifetime. In order to reduce the acquisition time, we implement an active memory reset that reduces $T_a$, thereby accelerating the convergence of the memory to the vacuum. In this section we describe how we  implement this reset mechanism.

We follow the same analysis as in Sec.~\ref{sec:parametric_pumping_ATS}, but we consider the term in $\Delta E_J$ that was previously neglected. At the flux saddle point, this term contributes a non-linear interaction of the form $U^\Delta(\bp)=2\Delta E_J\cos(\varepsilon_p\cos(\omega_p t))\cos(\bp)$. Assuming $\varepsilon_p, |\bp|\ll 1$, we obtain
\begin{equation}
\label{eq:UDelta}
U^\Delta(\bp)\approx2\Delta E_J(1-\frac{1}{2}\varepsilon_p^2\cos(\omega_p t)^2)(1-\frac{1}{2}\bp^2)\;.
\end{equation}
The amplitude of the reset pump is denoted $\varepsilon_p^\text{reset}$ and its  frequency is set to $\omega_p^\text{reset}=(\omega_b-\omega_a)/2$. Performing a rotating wave approximation on Eq.~\eqref{eq:UDelta}, we find
\begin{equation}
U^\Delta_{\omega_p=\omega_p^\text{reset}}(\bp)/\hbar=g_\text{reset}(\ba\bb^\dag+\ba^\dag\bb)\;,
\end{equation}
where $g_\text{reset}=\Delta E_J{\varepsilon_p^\text{reset}}^2\varphi_a\varphi_b/8\hbar$. We place ourselves in the regime where $g_\text{reset}\ll\kappa_b$, which induces a dissipation rate on the memory of the form $\kappa_a^\text{reset}=4g_\text{reset}^2/\kappa_b$.

In practice, we calibrate the reset pump by measuring the memory lifetime in the presence of the reset pump and vary its frequency (Fig.~\ref{fig:supmat_active_reset}). When the reset pump frequency is far detuned from the frequency matching condition $(\omega_b-\omega_a)/2$, the memory lifetime is unaffected by the pump, remaining around $20~\mu$s. However, when the pump frequency is set to $(\omega_b-\omega_a)/2$, the memory lifetime drops to $\approx 5~\mu$s. Note that the buffer and memory frequencies that enter this frequency matching condition are those in the presence of the reset pump.

\section{Photon number calibration}
\label{sec:photon_number_calibration}

\begin{figure}
\centering
\includegraphics[width=1\columnwidth]{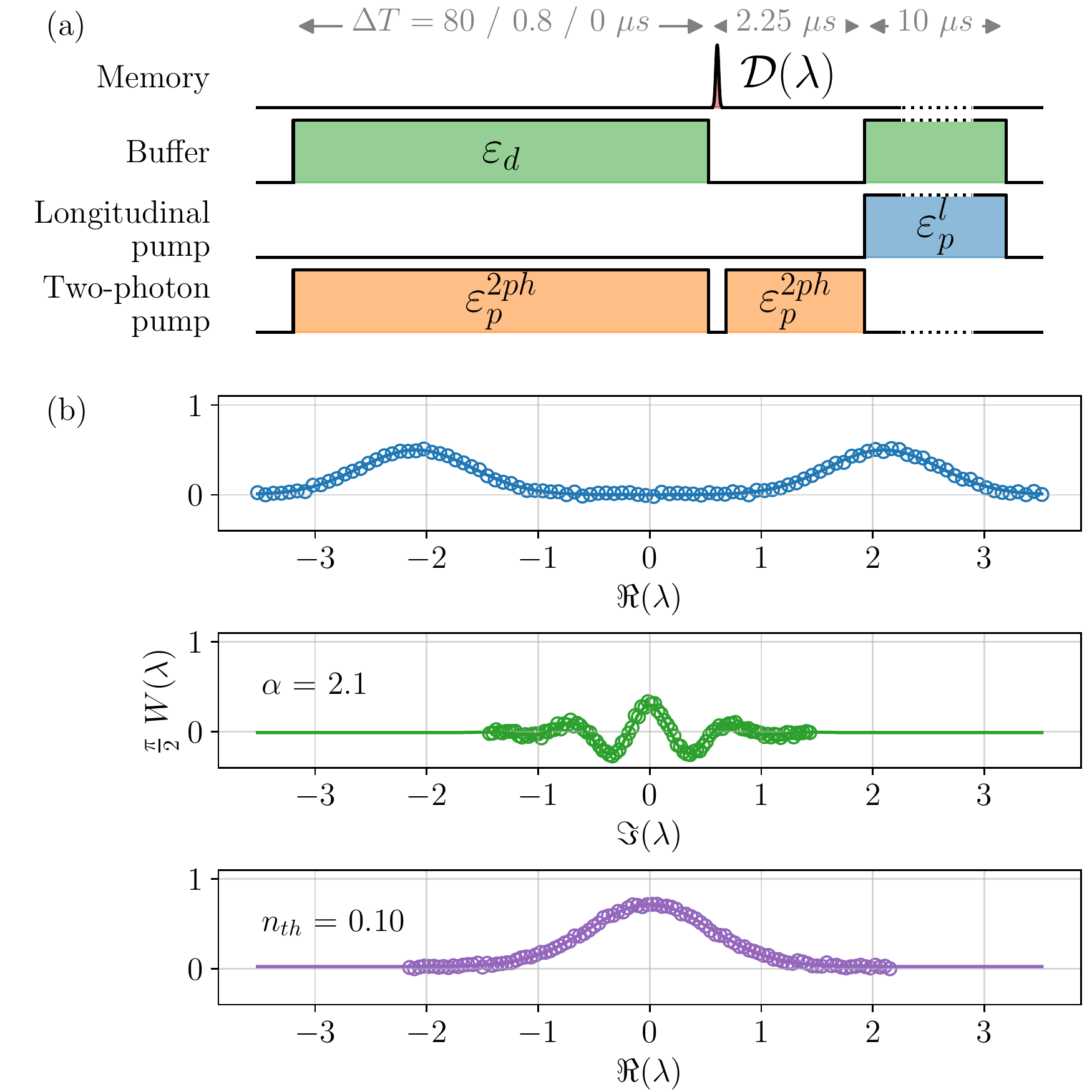}
\caption{{\bf Photon number calibration.} (a) Pulse sequence: a two-photon pump and buffer drive are active for $\Delta T=80, 0.8, 0~\mu$s, thereby preparing a statistical mixture of $\ket{\pm\alpha}$, a cat state $\ket{+}_\alpha$ and the thermal vacuum, respectively. Next, a Wigner tomography protocol is implemented (see text). (b) Cuts of the Wigner functions (y-axis) of these three states (top to bottom) along the real (top and bottom) and imaginary (middle) axis. The data (open circles) are overlaid with fits to the analytical formula displayed in Eq.~\eqref{eq:supmat_analytics} (solid lines).}
\label{fig:supmat_photon_number_calibration}
\end{figure}

The readout protocol based on the memory-buffer longitudinal coupling yields a buffer signal proportional to the number of photons in the memory. Hence the proportionality constant must be calibrated. Even though the experiment shown Fig.~\ref{fig:supmat_deflation} allows to calibrate it in principle, it is not very accurate. In order to access this crucial information, we elaborate a simple Wigner tomography tool. The signal to noise ratio of this method is lower than the one presented in the main text, however it requires less calibrations and is therefore more robust and easier to implement.

\paragraph{Wigner tomography:}
Remarkably, since two-photon dissipation (with a buffer drive set to zero) removes photons in pairs, it maps all even parity states to the vacuum $\ket{n=0}$, and all odd parity states to Fock state $\ket{n=1}$. This maps parity to photon number, which is measurable through longitudinal coupling to the buffer. A memory displacement of amplitude $\lambda$ followed by a two-photon pump applied without buffer drive during $2~\mu$s (Fig.~\ref{fig:supmat_deflation}) therefore measures the Wigner function of the memory's initial state $W(\lambda)$ (Fig.~\ref{sec:photon_number_calibration}a). Single-photon losses after mapping to 0/1 and the overlap between the histograms of the longitudinal readout corresponding to $\ket{n=0}$ and $\ket{n=1}$ lead to a global reduction of the scale of the Wigner tomography.

\paragraph{Calibration:} We apply this tomography protocol to three different initial state preparations: (i) applying the two-photon pump and buffer drive for $\Delta T=80~\mu$s resulting in an almost equal statistical mixture of $\ket{\pm}_\alpha$, (ii) applying the two-photon pump and buffer drive for $\Delta T=800$~ns resulting in a mixture biased towards $\ket{+}_\alpha$ and  (iii) no pump or drive at all, resulting in the vacuum state with some residual thermal population. Interestingly, note that while the thermal vacuum state displays a Wigner distribution that is broadened by thermal fluctuations, states (i) and (ii) are cooled into the cat-qubit manifold, and therefore the memory thermal occupation will impact the state reached within this manifold, but does not lead to thermal broadening. This asymmetry will be leveraged to calibrate the memory thermal occupation. Cuts of the Wigner functions of these three states are shown in Fig.~\ref{sec:photon_number_calibration}.b, and overlaid with their fits to analytic formula \cite{Haroche2006}:

\begin{eqnarray}
\label{eq:supmat_analytics}
W_{\Delta T=80~\mu\text{s}}(\lambda_r, 0)&=&{A}\left(e^{-2(B\lambda_r-\alpha)^2}+e^{-2(B\lambda_r+\alpha)^2}\right)\notag\\
W_{\Delta T=0.8~\mu\text{s}}(0,\lambda_i)&=&{C}e^{-2(B\lambda_i)^2}\left(e^{-2\alpha^2}+\cos(4\alpha B\lambda_i)\right)\notag\\
W_{\Delta T=0~\mu\text{s}}(\lambda_r,0)&=&{D}e^{-2(B\lambda_r)^2/(2n_{th,a}+1)}\;,
\end{eqnarray}
where $\lambda_{r,i}$ are the real and imaginary part of $\lambda$ and $A, B, C, D, \alpha, n_a^{th}$, as well as offsets, are free fit parameters. In essence, the frequency of the oscillations in (ii) sets an absolute scale $B$ for memory displacements in units of square root of photon number; the width of the Gaussian in (iii) sets the thermal population $n_a^{th}=10\%$ of the memory and the separation of the two Gaussians in (i) indicates the size of the cat-qubit $\alpha=2.1$. Furthermore, we do not observe a broadening of the two Gaussians in (i), which indicates a small thermal population  $n^b_{th}$ of the buffer. Finally, the constants $A,C,D$ encode the different fidelities of state preparation and measurement.

\section{Quantum Zeno dynamics}

\begin{figure*}
\centering
\includegraphics[width=1.2\columnwidth]{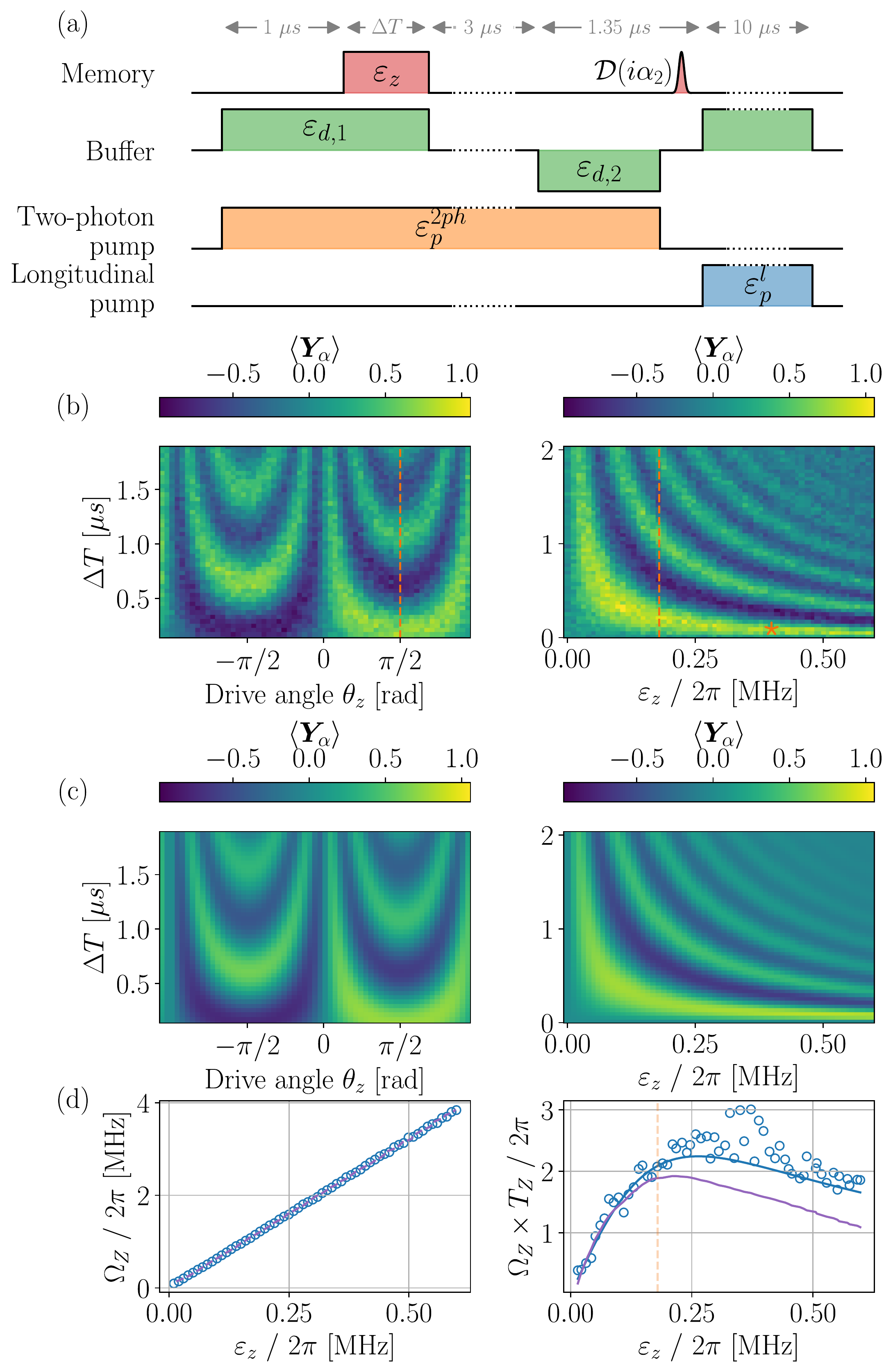}
\caption{{\bf Quantum Zeno dynamics.} (a) Pulse sequence: a two-photon pump and buffer drive prepare $\ket{+}_\alpha$ for $|\alpha^2|=2.6$. A memory drive is then activated with a variable phase and for a variable time $\Delta T$ inducing coherent rotations in the steady state manifold. The rest of the pulse sequence measures $\bY_\alpha$ as described in Fig.~2. (b-left) Buffer output proportional to the expectation value of $\bY_\alpha$ (color) as a function of the memory drive angle (x-axis) and duration (y-axis). The memory drive angle is set to maximize the oscillation frequency (orange dashed line). (b-right) With the angle now calibrated, the memory drive amplitude is varied (x-axis) and the expectation value of $\bY_\alpha$ (color) is measured versus pulse duration (y-axis). (c-left and c-right) Numerical simulation of the pulse sequence (a) with the circuit parameters given in Table \ref{table:supmat_qutip_params}. (d-left) From (b-right) we extract the oscillation frequency (y-axis) versus memory drive amplitude (x-axis). The data (open circles) follow the expected linear trend (solid line). We calibrate $\epsilon_z$ by fitting the data to the relation $\Omega_z=4|\alpha|\epsilon_z$ \cite{Guillaud2019}. ({d}-right) Number of coherent oscillations in units of oscillation frequency times fitted dephasing time (y-axis) versus memory drive amplitude $\varepsilon_Z$ (x-axis). The data (open circles) reach a maximum when $\varepsilon_Z$ balances rotation speed and induced dephasing. Surprisingly, the data are well explained by an analytical formula (Eq.~\eqref{eq:Gammaz}) valid in the adiabatic regime $8 g_2\alpha/\kappa_b\ll 1$ (solid blue line), although in our experiment, $8g_2\alpha/\kappa_b\sim 3.6$. On the other hand, the full master equation  simulation underestimates the dephasing time (solid purple line). The vertical orange dashed lines in (b-right) and (d-right) are placed at the memory drive amplitude picked for the data of Fig.~4. In an attempt to optimize the holonomic gate, we chose a stronger memory drive for a faster $\pi/2$ pulse marked by the orange star in (b-right).}
\label{fig:supmat_rabi_zeno}
\end{figure*}

One of the main results of this work is to perform coherent oscillations of superpositions of metastable states with macroscopic bit-flip times. These oscillations are induced by the interplay of two-photon dissipation and a coherent drive on the memory. While the drive attempts to translate the memory state in phase space, two-photon dissipation continuously stabilizes the manifold spanned by $\ket{\pm}_\alpha$. As a result, when the phase of memory drive is 90 degrees out of phase with $\alpha$, we observe coherent oscillations between $\ket{\pm}_\alpha$. The coherent motion in a manifold stabilized by dissipation is known as quantum Zeno dynamics \cite{Signoles2014, Schfer2014, Bretheau2015, Touzard2018}.

In this section we describe the procedure we follow to calibrate the amplitude and phase of the memory drive (Fig.~\ref{fig:supmat_rabi_zeno}). Starting from a memory in  vacuum, we implement the pulse sequence displayed in Fig.~\ref{fig:supmat_rabi_zeno}a, where the phase $\theta_Z$ and duration $\Delta T$ of the memory drive is varied. We observe $\langle Y_\alpha\rangle$ oscillating versus $\Delta T$, with an oscillation frequency that depends strongly on $\theta_Z$ (Fig.~\ref{fig:supmat_rabi_zeno}b). We identify the phase at witch the oscillation frequency is minimal with  $\theta_Z=0$~mod~$\pi$, that is, the memory is driven in phase with $\alpha$. On the other hand, we identify the phase at which the oscillation frequency of $\langle Y_\alpha\rangle$ is maximal with $\theta_Z=\pi/2$~mod~$\pi$, that is, the memory drive is 90 degrees out of phase with $\alpha$. To optimize the $Z$ gate speed, we set $\theta_Z=\pi/2$ in the pulse sequence of Fig.~2a. Surprisingly, when we measure the evolution of $Z_\alpha$ in the presence of a memory drive with $\theta_Z=\pi/2$, we find that the $Z_\alpha$ population is skewed towards one of the two pointer states $\ket{\pm\alpha}$. Therefore, for the data of Fig.~4a,b, we tune $\theta_Z$ in a $\pm15^{\circ}$ interval around $90^{\circ}$ to balance the population of $\ket{\pm\alpha}$. Understanding the origin of this deviation will be addressed in future work.

Having fixed the memory drive phase, we now vary its amplitude $\varepsilon_Z$. The data representing $\langle Y_\alpha\rangle$ as a function of drive amplitude and duration are displayed in Fig.~\ref{fig:supmat_rabi_zeno}c. Two features are visible. First, as the drive amplitude is increased, the frequency of the oscillation increases proportionally (Fig.~\ref{fig:supmat_rabi_zeno}d). Second, these oscillations decay faster as $\varepsilon_Z$ is increased, which is a manifestation of so-called non-adiabatic errors (Sec.~\ref{sec:gammaZ_epsilonZ}). We plot the number of oscillations per decay time in Fig.~\ref{fig:supmat_rabi_zeno}e, and find that there is an optimal gate amplitude that balances gate duration and induced dephasing.

\section{State preparation}
\begin{figure*}
\centering
\includegraphics[width=2\columnwidth]{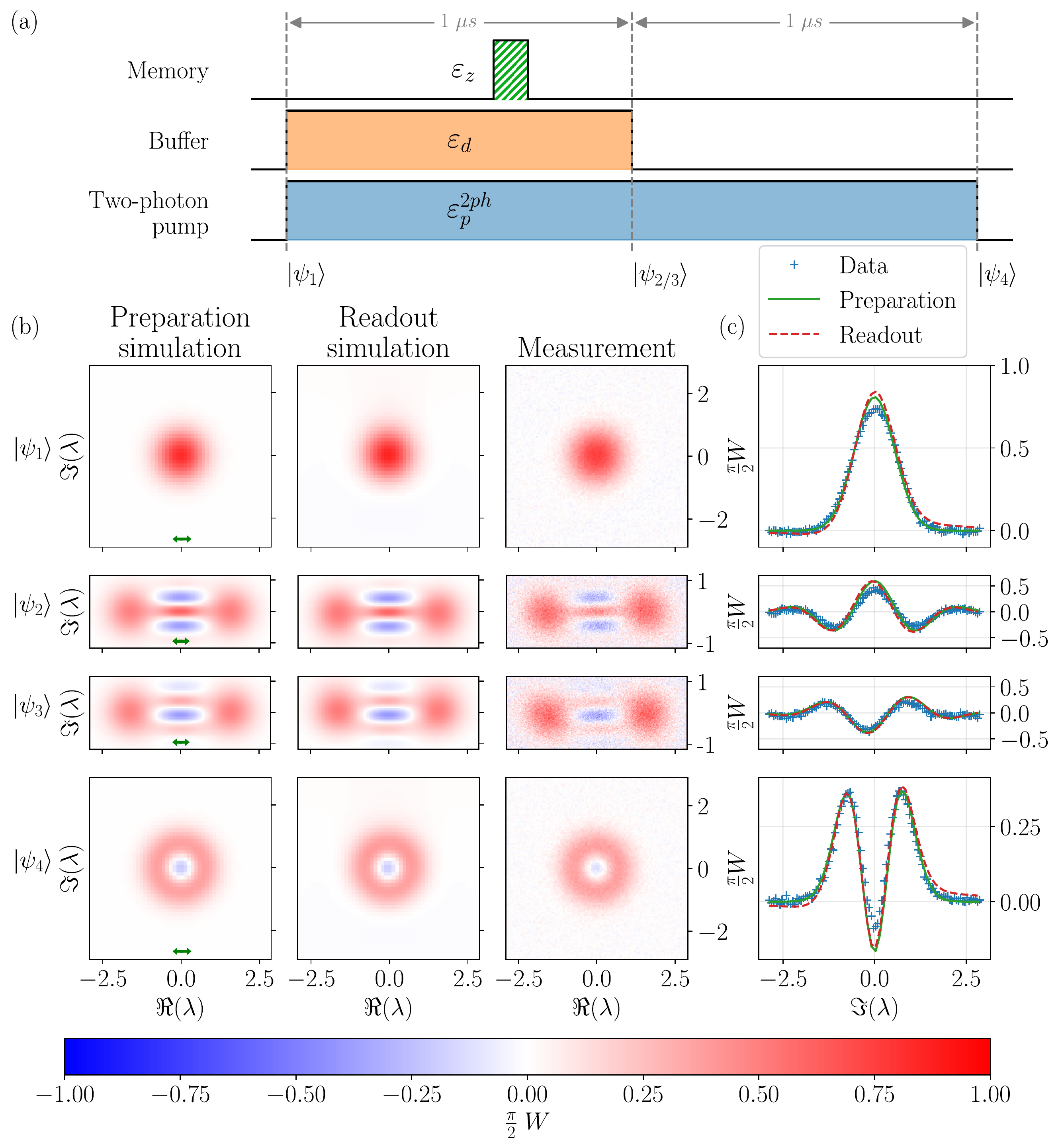}
\caption{{\bf Quantum state tomography.} We prepare and image states $\ket{0}, \ket{+}_\alpha, \ket{-}_\alpha$ and $\ket{1}$, labeled $\ket{\psi_1}$ to $\ket{\psi_4}$ for $\alpha^2=2.6$. (a) Pulse sequence: starting from the vacuum $\ket{\psi_1}$, the two-photon pump (blue) and buffer drive (orange) are activated. When no memory drive is played (dashed green), the even cat-state $\ket{\psi_2}$ is prepared. On the other hand, when the memory drive is played, it implements a $Z(\pi)$ gate thereby preparing the odd cat state $\ket{\psi_3}$. Finally, in the latter case, the buffer drive is turned off to arrive at the first Fock state $\ket{\psi_4}$. At each step of this pulse sequence, the protocol of Fig.~2 is implemented to acquire the Wigner function of these states. (b) Wigner amplitude (color) over the complex plane (x-y axis) for each state $\ket{\psi_{1,2,3,4}}$. The data ("Measurement" column) are well reproduced by the Lindblad simulation of the entire protocol including the preparation only ("Preparation simulation" column) and both preparation and tomography pulse sequences ("Readout simulation" column). (c) Integration of the Wigners in (b) over the interval marked by the green double arrow.}
\label{fig:supmat_states_preparation}
\end{figure*}

In Fig.~\ref{fig:supmat_states_preparation} we detail the protocol followed to prepare the states that are imaged in Fig.~2. Also, we numerically simulate the pulse sequence for generation and tomography of these states and compare the simulated and measured Wigner functions.

\section{Theory}
In this section, we derive a semi-classical model for the dynamics described by Eq.~(1) (Sec.~\ref{sec:semi_classical_model}).
We then fully characterize the steady states of this semi-classical model (Sec.~\ref{sec:steady_states}),
derive the linearized evolution in their neighbourhood
and compute the associated eigenvalues to extract the confinement rate (Sec.~\ref{sec:confinement_rate}). This semi-classical analysis is well suited for $|\alpha|\gtrsim 1$ where the steady state manifold is spanned by weakly overlapping coherent states. We then proceed to the derivation of an analytical expression for the confinement rate when $\alpha=0$ (Sec.~\ref{sec:confinement_rate0}). Finally, we model our measurement scheme of real-time memory trajectories through the buffer in Sec.~\ref{sec:memory_trajectories}.

Note that a similar analysis was performed in Ref.~\cite{Lescanne2020} in the regime where $|g_2|\ll \kappa_b$ and $|\alpha|\gtrsim 1$. Here we generalize the analysis to arbitrary $g_2$ and $\kappa_b$ and we treat the case $\alpha=0$.

\subsection{Semi-classical model}
\label{sec:semi_classical_model}
We start by writing the Hamiltonian of Eq.~(1) with the substitution $\varepsilon_d=\alpha^2g_2^*$: 
\begin{eqnarray}
    \bH_{2ph} &=& g_2^* (\ba^2 - \alpha^2) \bb^\dagger +\mathrm{h.c.}\;.
\label{eq:H_2ph}
\end{eqnarray}
Next, using the Hamiltonian in the form given above, we compute the quantum Langevin equations associated to Eq.~(1):
\begin{eqnarray}
\dot\ba&=&-2ig_2\ba^\dag\bb\notag\\
\dot\bb&=& -ig_2^*(\ba^2-\alpha^2)-\frac{\kappa_b}{2}\bb+\sqrt{\kappa_b}\bb_{in}\;,\label{eq:langevin}
\end{eqnarray}
where $\bb_{in}$ is the incoming bath field on the buffer. Note that since we assume for simplicity that $\kappa_a$ is strictly zero in this analysis, the bath field on the memory  does not enter this equation. The semi-classical approximation consists in replacing every operator in Eq.~\eqref{eq:langevin} by its expectation value. We denote $a=\langle\ba\rangle$ and $b=\langle\bb\rangle$. Moreover, since  $\langle\bb_\text{in}\rangle=0$, we find:
\begin{eqnarray}
\dot a&=&-2ig_2 a^*b\notag\\
\dot b&=& -ig_2^*(a^2-\alpha^2)-\frac{\kappa_b}{2}b\;.\label{eq:semiclassical}
\end{eqnarray}

\subsection{Steady states}
\label{sec:steady_states}
The dynamics described in Eq.~\eqref{eq:semiclassical} feature three fixed points :
\begin{equation}
(a,b)=(0,{2ig_2^*\alpha^2}/{\kappa_b})\;,(a,b)=(\alpha,0)\;,(a,b)=(-\alpha,0)\;.
\end{equation}

\subsubsection{Unstable equilibrium}
Let us compute the linearized dynamics around $(a,b)=(0,{2ig_2^*\alpha^2}/{\kappa_b})$.
To this end, we define $\delta a=a-0=a$ and $\delta b=b-2ig_2^*\alpha^2/{\kappa_b}$, 
and compute the dynamics of $\delta a, \delta b$ up to first order terms :
\begin{equation}
\label{eq__lin_za_0}
\left\{
    \begin{aligned}
    \delta \dot a &= \frac{4|g_2|^2}{\kappa_b} \alpha^2 \delta a^* \\
    \delta \dot b &= -\frac{\kappa_b}2 \delta b\;.
    \end{aligned}
    \right.
\end{equation}
We thus see that $\Re(\delta a)$ is unstable around this equilibrium.

\subsubsection{Stable equilibria}
Let us compute the linearized dynamics around $(a,b)=(\alpha,0)$;
the case of $a = -\alpha$ is similar.
We define $\delta a =a-\alpha$ and $\delta b=b-0=b$, and compute the dynamics of $(\delta a, \delta b)$ up to first order terms :
\begin{equation}
\label{eq__lin_za_alpha}
\left\{
    \begin{aligned}
    \delta \dot a &= -2ig_2\alpha^* \, \delta b \\
    \delta \dot b &= -2ig_2^*\alpha \delta a - \frac{\kappa_b}2 \, \delta b\;.
    \end{aligned}
    \right.
\end{equation}

Notice that the structure of \eqref{eq__lin_za_alpha} implies that we can decouple the dynamics
of the real and imaginary parts of $\delta a$ and $\delta b$.
We define the real quantities $x_{a,b}, y_{a,b}$ such that $\delta a = x_a + i y_a$ and $\delta b = x_b + i y_b$. For simplicity, we assume $g_2$ and $\alpha$ to be real.
We then have
\begin{equation}
\begin{bmatrix}
    \dot x_a \\
    \dot y_b \\
    \dot y_a \\
    \dot x_b \\
\end{bmatrix}
=
\begin{bmatrix}
    0 		& 2g_2\alpha 		& 0 		& 0 \\
    -2g_2\alpha	& -\frac{\kappa_b}2	& 0		& 0 \\
    0		& 0			& 0		& -2g_2\alpha \\
    0		& 0			& 2g_2\alpha	& -\frac{\kappa_b}2 
\end{bmatrix}
\begin{bmatrix}
    x_a \\
    y_b \\
    y_a \\
    x_b
\end{bmatrix}
\end{equation}

Since a matrix and its transpose have the same eigenvalues,
the eigenvalues associated to this linearized dynamics are the same as the eigenvalues of :
\begin{equation}
A =
\begin{bmatrix}
    0	& 2g_2\alpha \\
    -2g_2\alpha	& -\frac{\kappa_b}2
\end{bmatrix}\;.
\end{equation}
Notice that $\det(A) = 4g_2^2\alpha^2 > 0$ and $\trace(A) = -\frac{\kappa_b}2 < 0$
so that the real part of each eigenvalue of $A$ is always negative.
We can thus already conclude that the equilibria $(a = \pm\alpha, \, b = 0)$ are stable
for all values of $\alpha, \, \kappa_b, \, g_2$. 

\subsection{Confinement rate in the semi-classical picture}
\label{sec:confinement_rate}
In this section we derive an expression for the confinement rate $\kappa_\text{conf}$ at which states locally converge towards the stable steady state $(a,b)=(\alpha,0)$ (the case $(a,b)=(-\alpha,0)$ is similar). Let us denote the characteristic polynomial of the matrix $A$ by
\begin{equation*}
\chi_A(\lambda) = \det(\lambda I - A) = \lambda^2 + \frac{\kappa_b}2 \lambda + 4g_2^2\alpha^2\;.
\end{equation*}
Its discriminant is given by
\begin{equation*}
\Delta = \frac{\kappa_b^2}4 - 16 g_2^2 \alpha^2
= \frac{\kappa_b^2}4 \left( 1 - \left(\frac{8g_2\alpha}{\kappa_b} \right) ^2 \right)\;.
\end{equation*}
The critical value of the parameters (\emph{i.e.} the frontier between real and complex conjugate eigenvalues) is thus given by the relation
\begin{equation*}
\frac{8g_2\alpha}{\kappa_b}\biggr \rvert_\text{critical} = 1\;.
\end{equation*}

\subsubsection{{Overdamped} regime : $8g_2\alpha < \kappa_b$.}
In this case, the eigenvalues are real and given by
\begin{equation*}
\lambda_{\pm} = -\frac{\kappa_b}4
\left( 1 \pm \sqrt{ 1 - \left(\frac{8g_2\alpha}{\kappa_b} \right) ^2 } \right).
\end{equation*}
The local convergence rate $\kappa_\text{conf}$ is thus governed by the eigenvalue of least magnitude
\begin{equation*} \kappa_\text{conf}\biggr \rvert_\text{overdamped}=-2\lambda_- = \frac{\kappa_b}{2}
\left( 1 - \sqrt{ 1 - \left(\frac{8g_2\alpha}{\kappa_b} \right) ^2 } \right).
\end{equation*}
where the factor 2 comes from the fact that Eq.~\eqref{eq__lin_za_alpha} describes the amplitudes dynamics whereas the convergence rate should be homologous to an energy decay rate.
Note that in the limit $g_2\ll\kappa_b$,
a first order approximation of the above value is
\begin{equation*}
\kappa_\text{conf}\biggr \rvert_{g_2\ll\kappa_b}= \frac {16g_2^2\alpha^2}{\kappa_b} = 4 \kappa_2 \alpha^2
\end{equation*}
where $\kappa_2 = \frac{4g_2^2}{\kappa_b}$.
We recover the value of $\kappa_\text{conf}$ obtained in \cite{Lescanne2020}
by a semi-classical analysis of the effective one mode model when $g_2\ll\kappa_b$, up to a factor 2 due to the new convention.

\subsubsection{Underdamped regime : $8g_2\alpha > \kappa_b$.}
In this case, the eigenvalues are complex conjugate and given by
\[ \lambda_{\pm} = -\frac{\kappa_b}4
\left( 1 \pm i \sqrt{ \left(\frac{8g_2\alpha}{\kappa_b} \right) ^2 -1 } \right). \]
The local convergence rate is thus given by their shared real part
\begin{equation*}
\kappa_\text{conf}\biggr \rvert_\text{underdamped}=-2\Re(\lambda_{\pm}) = \frac{\kappa_b}2
\end{equation*}

\subsubsection{Regime of the experiment:} In our experiment, $\kappa_b/2\pi=2.6~$MHz and $g_2/2\pi=1$~MHz so $\kappa_b/8g_2\approx 0.3$. Hence for $\alpha>0.3$, we are in the underdamped regime, and $\kappa_\text{conf}=\kappa_b/2$ as announced in the main text. 

{This analysis is valid in the regime where $|\alpha|\gtrsim 1$ where the steady state manifold is spanned by weakly overlapping coherent states. In the next section we tackle the case where $\alpha=0$, a regime we enter during the execution of the holonomic gate.}

\subsection{Confinement rate for $\alpha=0$}
\label{sec:confinement_rate0}
{When $\alpha=0$, we define the confinement rate as $\kappa_\text{conf}=-2\Re(\lambda_\text{min})$, where $\lambda_\text{min}$ is the non-zero eigenvalue of the Liouvillian superoperator with the smallest real part in absolute value. A detailed mathematical analysis of the eigenvalues of the Liouvillian will appear in a future publication \cite{Sellem202X}. Here we provide a simplified analysis that leads us to an analytical expression of $\kappa_\text{conf}$.}

{Neglecting thermal excitations and memory losses ($n_a^{th}=n_b^{th}=0$, $\kappa_a=0$), we rewrite the Lindblad master equation \ref{eq:master_equation} in the superoperator formalism \cite{Albert2014}:}
{
\begin{align*}
    \frac{d\kket{\brho}}{dt} &= \hat{\mathcal{L}} \kket{\brho} \\
    \hat{\mathcal{L}} &= \mathbf{1} \otimes \bG^* + \bG \otimes \mathbf{1} + \kappa_b \bb \otimes \bb^* \\
    \bG &= -i g_2\left( \ba^2\bb^\dagger + {\ba^\dagger}^2 \bb \right) - \frac{\kappa_b}{2} \bb^\dagger \bb\;,
\end{align*}
}
{where $\kket{\brho}$ is the vectorized form of the density matrix $\brho$, $\hat{\mathcal{L}}$ is the Liouvillian superoperator, and $\mathbf{1}$ is the identity operator. For clarity, note that considering $N_a$ Fock states in the memory, and $N_b$ in the buffer, $\hat{\mathcal{L}}$ is a matrix of size $(N_a\times N_b)^2$ by $(N_a\times N_b)^2$, and $\kket{\brho}$ is a vector of size $(N_a\times N_b)^2$. Also, the tensor product is in between spaces that are each of dimension $N_a\times N_b$. Finally, the star symbol $^*$ refers to the complex conjugate, while as previously, the dagger symbol $^\dagger$ refers to the complex conjugate and transpose. Our simplified analysis consists in analyzing $\hat{\mathcal{L}}$ on the smallest set of states that captures the confining dynamics. One can immediately verify that $\ket{0,0}$ and $\ket{1,0}$ - left and right indices refer to the Fock state number in the memory and buffer respectively - are steady states of the dynamics. We denote the manifold of steady states $\mathcal{H}_0=\text{Span}(\ket{0,0}, \ket{1,0})$. We now consider exiting $\mathcal{H}_0$ by applying a memory excitation $\ba^\dagger$ on these states. This transfers $\ket{0,0}\rightarrow \ket{1,0}$ and $\ket{1,0}\rightarrow\ket{2,0}$. This last state lies outside the manifold of steady states, and will be transferred back through the combined action of $\ba^2\bb^\dagger$ that transfers $\ket{2,0}\rightarrow\ket{0,1}$ and buffer losses that transfers $\ket{0,1}\rightarrow \ket{0,0}$. Hence we consider the space $\mathcal{H}_1=\text{Span}(\ket{2,0}, \ket{0,1})$ and restrict ourselves to the study of $\hat{\mathcal{L}}$ over a space decomposed in the four following blocks : $\mathcal{H}=(\mathcal{H}_0\otimes\mathcal{H}_0) \oplus (\mathcal{H}_0\otimes\mathcal{H}_1) \oplus (\mathcal{H}_1\otimes\mathcal{H}_0) \oplus (\mathcal{H}_1\otimes\mathcal{H}_1)$. One can verify that extending the analysis to states accessed by applying $\bb^\dag$ to $\mathcal{H}_0$ does not change the value of $\lambda_\text{min}$.}

{Note that $\mathbf{G}\mathcal{H}_0=0$, $\mathbf{G}\mathcal{H}_1\subset\mathcal{H}_1$, $\bb\mathcal{H}_0=0$ and $\bb\mathcal{H}_1\subset \mathcal{H}_0$. Therefore, expressing the matrix of $\hat{\mathcal{L}}$ on $\mathcal{H}$ is block upper triangular. It's eigenvalues are those of its diagonal blocks, which are of the form $\lambda_1+\lambda_2^*$ where $\lambda_{1,2}$ are eigenvalues of $\mathbf{G}$. It then suffices to consider the matrix of $\mathbf{G}$ over $\mathcal{H}_1$:}
{\begin{equation*}
\mathbf{G}_1=\begin{bmatrix}
    0 & -i g_2 \sqrt{2} \\
    - i g_2 \sqrt{2} & -\frac{\kappa_b}{2}
\end{bmatrix}\;,
\end{equation*}}
{and its eigenvalues $\lambda_\pm$ :
\begin{equation*}
    \lambda_\pm = -\frac{\kappa_b}{4}\left( 1 \mp \sqrt{1 - 32\frac{{g_2}^2}{\kappa_b^2}} \right)\;.
\end{equation*}}
{
Finally, the confinement rate $\kappa_{conf}$ for $\alpha=0$ is:
\begin{equation*}
    \kappa_{conf}\biggr \rvert^{\alpha=0} = \frac{\kappa_b}{2}\Re\left( 1 - \sqrt{1-32\frac{{g_2}^2}{\kappa_b^2}} \right)\;.
\end{equation*}}
{Interestingly, in the overdamped regime $g_2\ll\kappa_b/\sqrt{32}$:
$$\kappa_{conf}\biggr\rvert^{\alpha=0}_\text{overdamped}=8g_2^2/\kappa_b\;,$$ 
and in the underdamped regime $g_2>\kappa_b/\sqrt{32}$, which is the regime of the experiment:
$$\kappa_{conf}\biggr \rvert^{\alpha=0}_\text{underdamped}=\kappa_b/2\;.$$}

\subsection{Memory trajectories}
\label{sec:memory_trajectories}
This section discusses the measurement technique employed in the main text to measure macroscopic bit-flip times. By weakly driving the memory mode, the system moves away from its steady state thus inducing a response on the buffer mode that contains information about the bit value $\pm\alpha$ of the memory state~\cite{Gautier2023}.

The Hamiltonian in the presence of the memory drive of amplitude  $\varepsilon_Z$ is $\bH_{2ph}+\bH_Z$ where $\bH_Z=\varepsilon_Z \ba^\dagger + \varepsilon_Z^*\ba$. Following the same steps leading to Eq.\eqref{eq__lin_za_alpha}, the local dynamics around $a=\pm\alpha$ and $b=0$ read: 
\begin{equation}
\label{eq:semiclassical_epsilonZ}
\left\{
    \begin{aligned}
    \delta \dot a &= \mp2ig_2\alpha^* \, \delta b-i\varepsilon_Z \\
    \delta \dot b &= \mp2ig_2^*\alpha \delta a - \frac{\kappa_b}2 \, \delta b\;.
    \end{aligned}
    \right.
\end{equation}
In steady state: $\delta\dot{a}=\delta\dot{b}=0$. Solving for $\delta b=b=\langle \bb \rangle$ in Eq.~\eqref{eq:semiclassical_epsilonZ}, we find
\begin{equation}
\label{eq:bpm}
\langle \bb \rangle=\mp \frac{\varepsilon_Z}{2\alpha^*g_2}\;.
\end{equation}

Therefore, by collecting the buffer output field, a weak measurement on the memory bit value is performed. Note that driving the system harder (i.e increasing $\varepsilon_Z$) would result in a higher readout contrast, but would also activate higher-order effects such as the ones neglected in this linearized dynamics that could eventually induce bit-flips by displacing the memory too far from its steady state.

\subsection{Drive-induced dephasing}
\label{sec:gammaZ_epsilonZ}
In the previous section, we have demonstrated that a memory drive leaks information about the memory state through the buffer output. Equivalently, this drive induces decoherence  - or dephasing - on the memory state. In this section, we compute the rate $\Gamma_Z^{\varepsilon_Z}$ of this drive-induced dephasing. From Eq.~\eqref{eq:bpm}, we find that depending on the state of the memory $\ket{\pm\alpha}$, the buffer converges to a coherent pointer state of amplitude $b_\pm=\mp \frac{\varepsilon_Z}{2\alpha^*g_2}$. The buffer then continuously radiates a field proportional to these amplitudes at a rate $\kappa_b/2$. Following \cite{Gambetta2006}, the induced dephasing is related to the rate at which information is gained on the memory state. Namely \cite{Gautier2023}:
\begin{eqnarray}
\label{eq:Gammaz}
\Gamma_Z^{\varepsilon_Z}&=&\frac{\kappa_b}{2}|b_+-b_-|^2\notag\\
&=&\frac{\kappa_b}{2}|\frac{\varepsilon_Z}{\alpha g_2}|^2\;.
\end{eqnarray}

\end{document}